%% file: paper.tex
\documentclass{article}
\usepackage[margin=1in]{geometry}

\title{Allocating with Priorities and Quotas: Algorithms, Complexity, and Dynamics}

\author{%
 Siddhartha Banerjee \\
 \texttt{sbanerjee@cornell.edu}\\
 Cornell University
 \and
 Matthew Eichhorn\\
 \texttt{mae226@cornell.edu}\\
 Cornell University
 \and
 David Kempe\\
 \texttt{david.m.kempe@gmail.com}\\
 University of Southern California%
}
\usepackage{booktabs} 

\usepackage{nicefrac}       
\usepackage{comment,bbm,graphicx}
\usepackage{booktabs} 
\usepackage{tcolorbox}

\usepackage{amsmath, amsthm, amssymb}

\usepackage{natbib}

\usepackage{parskip}

\usepackage{algorithm}
\usepackage[noend]{algorithmic}

\usepackage{enumitem,wrapfig,thm-restate}

\usepackage{xcolor}
\definecolor{myred}{HTML}{ea4335}
\definecolor{mygreen}{HTML}{41a756}
\definecolor{myblue}{HTML}{4285f4}

\usepackage[capitalize]{cleveref}
\crefname{lemma}{Lemma}{Lemmas}
\Crefname{lemma}{Lemma}{Lemmas}
\crefname{theorem}{Theorem}{Theorems}
\Crefname{theorem}{Theorem}{Theorems}

\usepackage{import}

\usepackage{tikz}
\newtheorem{theorem}{Theorem}
\newtheorem{proposition}{Proposition}
\newtheorem{corollary}{Corollary}
\newtheorem{lemma}{Lemma}

\newtheorem{definition}{Definition}

\DeclareMathOperator*{\argmax}{argmax}

\newcommand{\A}{\mathcal{A}}
\newcommand{\C}{\mathcal{C}}
\newcommand{\E}{\mathcal{E}}
\newcommand{\x}{\mathbf{x}}
\newcommand{\y}{\mathbf{y}}
\newcommand{\z}{\mathbf{z}}
\newcommand{\N}{\mathbf{N}}
\newcommand{\p}{\mathbf{p}}
\newcommand{\q}{\mathbf{q}}

\newcommand{\outside}{\perp}

\newcommand{\priority}{PR}
\newcommand{\quota}{QR}
\newcommand{\eligibility}{ER}
\newcommand{\pareto}{PE}
\newcommand{\stability}{CS}

\newcommand{\ax}[1]{\textbf{[#1]}}

\newcommand{\pmin}{p_{\textrm{min}}}

\usepackage{colortbl}
\usepackage{blkarray}

\begin{document}

\maketitle

\begin{abstract}
    \noindent In many applications such as rationing medical care and supplies, university admissions, and the assignment of public housing, the decision of who receives an allocation can be justified by various normative criteria (ethical, financial, legal, etc.). Such settings have motivated the following \emph{priority-respecting allocation problem}: several categories, each with a quota of interchangeable items, wish to allocate the items among a set of agents. Each category has a list of eligible agents and a priority ordering over these agents; agents may be eligible in multiple categories. The goal is to select a \emph{valid} allocation: one that respects quotas, eligibility, and priorities and ensures Pareto efficiency. \\[-6pt]

    \noindent We provide a complete algorithmic characterization of all valid allocations, exhibiting a bijection between sets of agents who can be allocated and maximum-weight matchings under carefully chosen rank-based weights. While prior work provides a polynomial-time algorithm to locate a valid allocation, our characterization admits a simpler algorithm that enables two wide-reaching extensions:\\[-6pt]

    \noindent 1. Selecting valid allocations that satisfy additional criteria: Via three examples --- inclusion/exclusion of some chosen agent; agent-side Pareto efficiency vs.~welfare maximization; and fairness from the perspective of allocated vs.~unallocated agents --- we show that finding priority-respecting allocations subject to some secondary constraint straddles a complexity knife-edge; in each example, one problem variant can be solved efficiently, while a closely related variant is \textsf{NP}-hard.\\[-6pt]

    \noindent 2. Efficiency-envy tradeoffs in dynamic allocation: In settings where allocations must be made to $T$ agents arriving sequentially via some stochastic process, we show that while insisting on zero priority violations leads to an $\Omega(T)$ loss in efficiency, one can design allocation policies ensuring that the \emph{sum} of the efficiency loss and priority violations in hindsight is $O(1)$ (under mild regularity conditions on the arrival process).
\end{abstract}

\import{./}{s1_intro.tex}

\import{./}{s2_model.tex}
\import{./}{s3_alg.tex}

\import{./}{s4a_hardness.tex}
\import{./}{s4b_utility.tex}
\import{./}{s4c_threshold.tex}
\import{./}{s5_online.tex}
\import{./}{s6_conclusion.tex}

\newpage
\section*{Acknowledgements}
The authors gratefully acknowledge support from AFOSR grant FA9550-23-1-0068, ARO MURI grant W911NF-19-1-0217, NSF grants ECCS-1847393 and CNS-195599, and the Simons Institute for the Theory of Computing. The authors also thank Oktay G\"{u}nl\"{u}k, Karola M\'{e}sz\'{a}ros, Rakesh Vohra, and the participants at the 2022 ACM Symposium on Foundations of Responsible Computing (FORC) for useful comments that helped shape this paper.

\bibliographystyle{abbrvnat}
\bibliography{refs}

\newpage
\appendix
\import{./}{a_scarf.tex}

\end{document}

%% file: s1_intro.tex
\section{Introduction} 
\label{sec:intro}

A core socio-economic question is how to ration scarce resources without money.
While not new, this question has been forcefully reintroduced into public consciousness by COVID-19~\citep{white2020framework,andrews2021no,emanuel2020fair,binkley2020ethical,pathak2021fair}. Defining ``good'' allocations is far from straightforward, as legal, financial, and ethical considerations can lead to nuanced, often clashing, requirements. For example, consider the following:

\smallskip
\noindent\emph{Academic Fellowships}: Donors often define qualification requirements for named scholarships to promote students with certain demographics/backgrounds/skills.

\noindent\emph{Medical Care}: The COVAX program set standards for the equitable distribution of vaccines in developing countries, prioritizing vaccination of groups such as healthcare workers, the elderly, and individuals with comorbidities~\citep{covax}. 

\noindent\emph{Primary School Enrollment}: In Boston, half of a school's seats are reserved for students in the neighborhood, and priority is given to siblings~\citep{abdulkadirouglu2005boston}. Chicago requires that schools allocate roughly $25\%$ of seats to each of four socio-economic tiers~\citep{benabbou2019fairness}. Chile's School Inclusion Law defines which factors can/cannot be used to prioritize students, and has quotas for students with economic hardships~\citep{correa2021school}.
    
\noindent\emph{Public Housing}: Singapore's 1989 Ethnic Integration Policy places quotas on the number of public housing units that may be allocated to each of three major ethnic groups~\citep{benabbou2018diversity}.
\smallskip

\noindent The above settings broadly share the following features: a resource (scholarships, vaccines, school seats, housing) must be rationed among agents, whose number typically exceeds the available resource budget. The budget is split into several \emph{categories}, each of which has a \emph{quota} the category is responsible for distributing --- this is sometimes due to physical constraints (different schools/housing projects), and at other times to implement some social norm (fellowship funds reserved for local/international/under-represented students; vaccine quotas for countries/states/target populations). Each category has rules to determine which agents are \emph{eligible} for allocation.
Each agent wants up to a single unit of the resource
, but is indifferent as to which category allocates that unit\footnote{This may not hold in all settings --- for example, families do have preferences between schools and housing units. We return to this issue in~\cref{sec:utility}. Nevertheless, it is true up to first order that agents prefer being allocated to staying unallocated.}. Agents may be eligible in multiple categories, so categories must coordinate to maximize allocations. 
Finally, categories often define rankings (or \emph{priorities}) over eligible agents, which are intended to help choose (and justify) which eligible agents get allocated.
These rankings are often idiosyncratic, so there may be no natural way to compare agents across categories.

To understand how the above features (quotas, eligibility, priorities) restrict allocations, we build on the framework introduced by~\citet{pathak2021fair}, which has led to a line of work aiming to understand its properties~\cite{delacretaz2021processing,aziz2021efficient,biro2021complexity}. 
We briefly summarize the framework below; see~\cref{sec:model} for a formal model.
\begin{tcolorbox}
\noindent\textbf{The Priority-Respecting Allocation Problem}
\begin{itemize}[nosep,leftmargin=*]
\item[---] $q$ resource units, split into quotas $q_c$ for categories $c\in\C$, must be rationed to agents $\A$.
    \item[---] Each category $c$ has a set $\E_c\subseteq\A$ of \emph{eligible} agents, and a priority order $\succeq_c$ over $\E_c$.
    
    \item[---] Agent $a$ is allocated $x_{a,c}$ from each category $c$, with $\sum_{c\in\C}x_{a,c}\leq 1$. \hfill (\emph{unit demand})
    
    \item[---] Category $c$ can allocate only to agents in $\E_c$. \hfill (\emph{eligibility-respecting})
    
    \item[---] Category $c$ can allocate up to its quota, i.e.,  $\sum_{a\in\A}x_{a,c}\leq q_c$. \hfill (\emph{quota-respecting})
    
    \item[---] Category $c$ can allocate to agent $a$ only once all higher priority agents are allocated\\ i.e., $x_{ac} > 0 \implies \sum_{c'\in\C} x_{a',c'}=1$ for all agents $a'\succ_c a$. \hfill (\emph{priority-respecting})
\end{itemize}
\end{tcolorbox}
The above problem tries to formalize what policymakers desire when using quotas, eligibility rules, and priorities, by providing a test for determining whether an allocation ``respects'' these requirements or not.
The first two conditions impose that a category should only allocate from its quota, and only to eligible agents --- these are standard and easily implemented. 
The third condition interprets priorities as a requirement that a category never allocate to an agent if a higher-priority agent has not been satisfied. This requirement is trickier to implement (and verify) since the higher-priority agents may receive allocations from any category.
Nevertheless, the axioms are easy to satisfy --- for example, each category can sequentially pick the highest-priority unallocated agent(s) in its eligibility list (more generally, via \emph{serial dictatorship}; see~\cref{ssec:characterization}).

One issue with the above requirements, however, is that they do not consider the ``efficiency'' of an allocation. For example, allocating to no one satisfies all the requirements. More problematic are settings with partial eligibility, where even if each category allocates maximally (i.e., until it exhausts its quota or eligibility list), the allocation may still end up wasting resources (for example, see~\cref{fig:axioms}). 
A natural additional requirement, therefore, is for allocations to be \emph{Pareto efficient} --- whereby there is no way for agents to exchange allocations such that at least one agent ends up gaining while no one is worse off.
Ensuring Pareto efficiency in addition to the above requirements, however, seems challenging, and prior work~\citep{pathak2021fair,delacretaz2021processing,abdulkadirouglu2021priority} states this as an open question. More recently,~\citet{aziz2021efficient} provide a scheme for finding a particular maximum-size (and hence Pareto efficient) allocation by solving $|\A|$ bipartite matching problems ---  this result, however, does not give any insight into general Pareto efficient valid allocations, and/or how one can select from among such allocations to satisfy some secondary objective. \cref{fig:axioms,fig:non-convex} give some intuition into the challenge of finding priority-respecting allocation --- in particular,~\cref{fig:non-convex} shows that unlike maximum matchings, the set of priority-respecting allocations is not necessarily convex.

\subsection{Our Contributions}

Our work aims to characterize the set of \emph{valid allocations}: those which respect {eligibility, quotas, and priorities}, and are {Pareto efficient} (see~\cref{sec:model}). Our main result is paraphrased as follows:

\begin{quote}
    \emph{A set of agents can be allocated via a valid allocation if (\cref{thm:delta_perturb}) and only if (\cref{thm:realizability}) they are allocated under the maximum matching for a weighted matching instance with edge weights picked from a certain valid set.}
\end{quote}

The set of valid weights (\cref{def:perturb}) is based on perturbing\footnote{Importantly, the perturbations can be \emph{local} (i.e., the edge weight between agent $a$ and category $c$ only depends on $a$'s position in $c$'s priority order); this is surprising, as we show that no such result is possible for the set of \emph{stable} matchings. We present a discussion that connects the stable matching and priority respecting allocation problems as special cases of Scarf's Lemma in~\cref{sec:scarf}.} 
 the unweighted matching objective such that the perturbations are consistent with the priorities, and the total perturbation is small (at most $\nicefrac{1}{2}$).
As an immediate consequence, we get that \emph{every} valid allocation allocates the same number of units, which moreover equals the size of an optimal matching \emph{without} priority requirements. We also show that although the set of valid allocations is non-convex, every fractional valid allocation is realized as a convex combination of integral valid allocations (\cref{prop:fractionalvalid}).

More importantly, our transformation of the problem of locating valid allocations into cardinal welfare maximization enables two wide-reaching and practical extensions. 
First, in \cref{sec:examples}, we consider how to select valid allocations satisfying additional criteria through three case studies:
\begin{enumerate}
\item \textbf{Valid allocations with agent inclusion/exclusion}: In~\cref{sec:serviceable}, given an agent $a\in\A$, we ask if one can find a valid allocation $\x$ that \emph{excludes} $a$ (i.e., $\sum_{c\in\C}x_{a,c}=0$), or \emph{includes} $a$ (i.e., $\sum_{c\in\C}x_{a,c}>0$). We show that while the former problem can be efficiently solved, the latter is \textsf{NP}-hard (\cref{prop:servicible_hard}).
These results give a glimpse into the strange algorithmic landscape of valid allocations; note that both of these problems can be efficiently handled for maximum matchings and stable matchings (via the LP characterization of~\citet{vate1989linear}). 

\item \textbf{Incorporating agent preferences}: In~\cref{sec:utility}, we augment the basic priority-respecting allocation problem by incorporating agent preferences for categories. We show how to efficiently find allocations that respect eligibility, quotas and priorities, and are also Pareto efficient under the agents' preference orders (\cref{thm:utility_pe_2step}). 
On the other hand, we show that the problem of selecting a valid allocation that maximizes practically any aggregate function of agents' utilities is \textsf{NP}-hard (\cref{thm:max_utility_hard}). 
\item \textbf{Inner/outer allocation thresholds}: In \Cref{sec:thresholds}, we consider the problem of selecting valid allocations that optimize some score based on the \textit{inner allocation threshold} (the lowest-priority agent \emph{allocated} in each category) or, alternately, the \textit{outer allocation threshold} (the highest-priority agent in each category who remains \emph{unallocated}). Understanding these thresholds is important for auditing the ``fairness'' of an allocation.
We show that optimizing over inner thresholds can be done efficiently (\cref{prop:innerthrsum,prop:innerthrminmax}),  while optimizing over outer thresholds is \textsf{NP}-hard (\cref{prop:outerthrmaxmin,prop:outerthrsum}).
\end{enumerate}

Each of these cases demonstrates that selecting valid allocations straddles the line of computational efficiency; one possible variant of each case study admits an efficient algorithm (based on our main approach in \cref{sec:characterization}), while a closely related variant is computationally hard. 

\medskip
\noindent\textbf{Online Priority-Respecting Allocations with Dynamic Arrivals}: Finally, we consider an online variant of the priority-respecting allocation problem. 
Ours appears to be the first work that develops algorithms and performance guarantees for online settings, even though several of the original motivations for the priority-respecting allocation model were intrinsically tied to online allocation (of vaccines/medical supplies)~\cite{pathak2021fair}. 
In the setting we consider, agents belong to one of a small number of observable types. Agents arrive one at a time over $T$ rounds via some (known) underlying stochastic process, and the principal, after observing each agent's type, must immediately and irrevocably decide to either allocate this agent a unit from some category or leave the agent unallocated.
We demonstrate that completely forbidding priority violations leads to $\Omega(T)$ regret with all but exponentially small probability. However, by incorporating priority violations into the objective, our LP formulation of the problem enables the development of a Bayes selector online algorithm for which the sum of the expected efficiency loss and priority violations in hindsight is $O(1)$ (i.e., independent of the number of arriving agents and the resource budgets, but depending polynomially on the number of types and categories).

\subsection{Related Work}
\label{sec:related}

As mentioned, we build on the framework of~\citet{pathak2021fair}, which has inspired several follow-up papers. \citet{delacretaz2021processing} notes that since the axioms do not uniquely identify an allocation, different choices can induce biases; to allay this, he introduces a waterfilling-style \emph{simultaneous allocation} procedure that leads to a unique (fractional) outcome. On the other hand,~\citet{aziz2021efficient} introduce a 
procedure that results in a maximum-size allocation. 
\citet{aziz2021multi} describe how to incorporate diversity goals as an optimization objective in these settings. Finally, \citet{abdulkadirouglu2021priority} consider lower bounds on categories, and develop a choice rule that minimizes the number of priority violations in this setting.

A closely related problem to reserve allocation is fair division, where agents have preferences over (non-identical) items, and we seek a Pareto efficient division. The key distinction between these problems is that in fair division, agents' preferences determine the stability of an allocation, while in our setting, the justification for an allocation is dictated by category preferences, while its utility may depend on agent preferences. Nevertheless, the structures of desired allocations in both turn out to be quite similar. Our results provide some intuition as to why this is the case, as when viewed as an ordinal welfare maximization problem, it is clear that the two sides of the market are symmetric. Consequently, our techniques and results share commonalities with this literature. For example, our case study in~\cref{sec:serviceable} recovers results of~\citet{saban2015complexity} on computing match probabilities under random serial dictatorship. On the other hand, our perturbation approach is foreshadowed by~\citet{biro2021complexity}'s, who propose using (pseudo)welfare maximization for computing Pareto efficient fair division solutions. 

Finally, settings with two-sided preferences have a long history, stemming from Gale and Shapley's seminal work on the deferred acceptance (DA) algorithm~\citep{gale1962college}. While a fairly robust algorithm, DA can fail to compute a Pareto efficient allocation in the case of indifferences, as pointed out by~\citet{erdil2017two}. They describe an iterative procedure to Pareto improve an allocation while preserving its stability, illustrating that notions of stability and efficiency can be simultaneously realized. The flow-augmentation ideas in their improvement procedure share commonalities with our arguments in~\cref{sec:characterization}.

%% file: s2_model.tex
\section{Model} 
\label{sec:model}

\noindent\textbf{Resources, Categories, Quotas and Agents}:
A set $\A$ of \emph{agents} compete for $q$ units of a resource. The units are distributed to a set $\C$ of \emph{categories}, through which they are allocated. Each category $c \in \mathcal{C}$ is given an integer \emph{quota} of $q_c$ units to allocate, such that $q = \sum_c q_c$.

Each agent is unit-demand, i.e., can consume at most one unit of the resource. For the initial part, we assume that agents are indifferent as to which category provides their allocation; in~\cref{sec:utility}, we discuss how to incorporate agent utilities in this setting.

\medskip
\noindent\textbf{Eligibility and Priorities}:
Each category partitions $\A$ into a set of \emph{eligible} and \emph{ineligible} agents. The eligible agents are further partitioned into \emph{priority} tiers.

Formally, each category $c \in \C$ has an associated \emph{eligible set} of agents $\E_c \subseteq \A$, and a total preorder $\succeq_c$ over $\E_c$. 
Given any two agents $a,a' \in \A$, $a \succeq_c a'$ denotes that $a$ has weakly higher priority than $a'$ in $c$.
We write $a \succ_c a'$ when $a \succeq_c a'$ and $a' \not \succeq_c a$, so $a$ has (strictly) higher priority in $c$. 
Given any agent $a$ and any category $c$, we define the \emph{rank} of $a$ in $c$, denoted by $r_c(a)$, to be the length $\ell$ of the longest chain $a_1 \succ_c a_2 \succ_c \cdots \succ_c a_\ell = a$ with each $a_i \in \A$. Note that $1 \leq r_c(a) \leq |\A|$. 
We visualize instances using charts in the style of~\cref{fig:axioms}.

\medskip
\noindent\textbf{Desiderata for Valid Allocations}:
Our goal is to find allocations that respect eligibility, quotas, and priorities. Formally, a (fractional) allocation is a function $\x: \A \times \C \to [0,1]$, with $\sum_c x_{a,c} \leq 1$ for each $a \in \A$ (since agents are unit-demand). If all $x_{a,c} \in \{0,1\}$ (i.e., the matching is integral), then $\x$ coincides with an allocation map $\varphi: \A \to \C \cup \{\outside\}$ assigning each agent to either a category (through which they are allocated) or the outside option $\outside$ (if they remain unallocated). 

Moreover, the allocation must satisfy the three desiderata given below. These were proposed for integral allocations by~\citet{pathak2021fair}; we state the generalization for fractional matchings due to~\citet{delacretaz2021processing} since they naturally specialize to the integral case.

\begin{tcolorbox}
\begin{description}
    \item[{[\quota]} Quota Respecting:] 
    No category allocates more units than its quota.
    \[
        \sum_{a \in \A} x_{a,c} \leq q_c\quad \text{ for all } c\in\C.
    \]
    \item[{[\eligibility]} Eligibility Respecting:] 
    Agents are allocated only through eligible categories.
    \[
        x_{a,c} = 0 \quad \text{ for all }  c\in\C, a \notin \E_c.
    \]
    \item[{[\priority]} Priority Respecting:] 
    An agent receives any allocation through category $c$ only after all higher-priority agents in $c$ are \emph{fully} allocated.
    \[
        x_{a',c} > 0 \:\land\: a \succ_c a' \implies \sum_{c' \in \C} x_{a,c'} = 1 \quad \text{ for all } a,a'\in\A,c\in\C.
    \]
\end{description}
\end{tcolorbox}
    
While the above desiderata determine which allocations are invalid due to violating the prescribed properties, they still admit many allocations that are undesirable. In particular, setting all $x_{a,c}=0$ satisfies the preceding desiderata. 
A natural additional property is that any chosen allocation be \emph{Pareto efficient}, which we formalize as follows.
\begin{tcolorbox}
\begin{description}
    \item[{[\pareto]} Pareto Efficient:] 
    An allocation $\x$ satisfying \ax{\quota}, \ax{\eligibility}, and \ax{\priority} is Pareto efficient if there is no other allocation $\y$ satisfying these desiderata in which one agent gets a strictly greater allocation and no one receives a smaller allocation. Formally, for every $\y$ satisfying \ax{\quota},\ax{\eligibility}, and \ax{\priority},
    \[
        \text{there is an } a \in \A \; : \sum_{c \in \C} y_{a,c} > \sum_{c \in \C} x_{a,c} \implies \text{ there is an } \; a' \in \A \; : \sum_{c \in \C} y_{a',c} < \sum_{c \in \C} x_{a',c}.
    \]
\end{description}
\end{tcolorbox}

\begin{figure}[!t]
    \centering
    \begin{minipage}{0.24\textwidth}
        \begin{center}
        Allocation 1 \medskip
        
        \begin{tabular}{c|c|c}
             $\alpha$ {\scriptsize (1)} & $\beta$ {\scriptsize (1)} & $\gamma$ {\scriptsize (1)} \\ \hline
             \boxed{c} & \boxed{a} & b\,,\,{c} \\
             & {b} & a\,,\,\boxed{d}
        \end{tabular}
        \end{center}
    \end{minipage}
    \hfill
    \begin{minipage}{0.24\textwidth}
        \begin{center}
        Allocation 2 \medskip
        
        \begin{tabular}{c|c|c}
             $\alpha$ {\scriptsize (1)} & $\beta$ {\scriptsize (1)} & $\gamma$ {\scriptsize (1)} \\ \hline
             c & \boxed{a} & b\,,\,\boxed{c} \\
             & b & a\,,\,d
        \end{tabular}
        \end{center}
    \end{minipage}
    \hfill
     \begin{minipage}{0.24\textwidth}
        \begin{center}
        Allocation 3 \medskip
        
        \begin{tabular}{c|c|c}
             $\alpha$ {\scriptsize (1)} & $\beta$ {\scriptsize (1)} & $\gamma$ {\scriptsize (1)} \\ \hline
             \boxed{c} & \boxed{a} & \boxed{b}\,,\,c \\
             & b & a\,,\,d
        \end{tabular}
        \end{center}
    \end{minipage}
    \hfill
    \begin{minipage}{0.24\textwidth}
        \begin{center}
        Allocation 4 \medskip
        
        \begin{tabular}{c|c|c}
             $\alpha$ {\scriptsize (1)} & $\beta$ {\scriptsize (1)} & $\gamma$ {\scriptsize (1)} \\ \hline
             \boxed{c} & a & b\,,\,c \\
             & \boxed{b} & \boxed{a}\,,\,d
        \end{tabular}
        \end{center}
    \end{minipage}
    \caption{\small\it Four (integer) allocations in an instance with $\C = \{\alpha,\beta,\gamma\}$ with quotas $(1,1,1)$, and $\A = \{a,b,c,d\}$. Each category $c$ lists its eligible agents $\E_c$, where the agents listed in the $i$'th row have rank $r_c(\cdot) = i$. \\
    --- Allocation 1 violates \emph{\ax{\priority}}: $d$ is allocated in category $\gamma$, but $b$, who has higher priority, remains unallocated.\\ 
    --- Allocation 2 violates \emph{\ax{\pareto}}, as it is Pareto dominated by Allocation 3; note, however, that it is non-wasteful.\\ 
    --- Allocations 3 and 4 are both valid (\cref{def:valid}) and allocate to the same set of agents.\\
    --- Allocation 4 violates \emph{\ax{\stability}} (\cref{ssec:characterization}), as $\beta$ and $\gamma$ can swap to allocate to higher-priority agents.}
    \label{fig:axioms}
\end{figure}

As a special case of this definition, an \emph{integral} allocation is Pareto efficient if and only if there is no feasible way to allocate to a strict superset of agents. Although quite natural, Pareto efficiency has not been directly addressed in previous work. \citet{pathak2021fair} and \citet{delacretaz2021processing} consider a weaker \emph{non-wastefulness} property that stipulates that in any category with unallocated quota, all eligible agents must be fully allocated. It is easy to construct settings that admit non-wasteful but Pareto inefficient allocations; for example, see the Allocation 2 in~\cref{fig:axioms}.
\citet{aziz2021efficient} strengthen non-wastefulness to a \emph{maximality} property: any selected valid allocation must maximize the total number of allocated units. While maximality clearly implies Pareto efficiency, we show in~\cref{prop:pe_is_max} that the two properties are in fact equivalent in this setting. 
We find Pareto efficiency to be more natural, both in this desideratum and when extending to settings with agent preferences. Together, our four desiderata provide a notion of a \emph{valid} allocation: 

\begin{definition}[\bf Valid Allocation]    
    \label{def:valid}
    An allocation is \emph{valid} if it satisfies \emph{\ax{\quota}}, \emph{\ax{\eligibility}}, \emph{\ax{\priority}}, and \emph{\ax{\pareto}}.
\end{definition}

As an illustration, \cref{fig:axioms} depicts possible allocations for the same instance: while the first violates \ax{\priority} and the second violates \ax{\pareto}, the third and fourth are both valid allocations that, moreover, allocate to the same set of agents $\{a,b,c\}$. Note also that $\{a,b,c\}$ is the only set of agents who can be allocated via a valid allocation; there is no valid allocation in which agent $d$ gets allocated. (We revisit this idea in~\cref{sec:serviceable}.)

%% file: s3_alg.tex
\section{An Algorithmic Characterization of Valid Allocations}
\label{sec:characterization}

The primary concern of our work is to develop efficient algorithms that, given an instance with quotas, eligibility lists, and priorities, select a valid allocation satisfying some additional properties. 
To this end, we require an algorithmic way to characterize the set of valid allocations.
Note that it is straightforward to find an allocation that satisfies \ax{\quota}, \ax{\eligibility}, and \ax{\priority} (and is also non-wasteful \cite{pathak2021fair,delacretaz2021processing})  --- for example, via a round-robin policy where each category sequentially picks their top remaining agent. 
However, as Allocation 2 in~\cref{fig:axioms} demonstrates, this may not ensure \ax{\pareto}.
On the other hand, any maximum-cardinality matching satisfies \ax{\quota} and \ax{\pareto}. The challenge is to achieve all four desiderata simultaneously. 

Our main result shows that there is in fact a \emph{bijection between the set of agents selected in valid allocations, and the set of maximum weight matchings under certain valid weights}. In \cref{ssec:lp}, we show that in any instance, a valid allocation can be found using a simple weighted bipartite matching LP. Subsequently, in~\cref{ssec:characterization}, we explore some consequences of this LP formulation, including a complete characterization of all valid allocations and a discussion of their geometry; we also show that such a characterization is impossible for stable matchings.

\subsection{Finding Valid Allocations via Weighted Matchings} 
\label{ssec:lp}

As the basis of our formulation, we start with the following LP, which we denote by $(P_0)$.


\begin{align*}
    (P_0) \hspace{40pt} & \textrm{max} & V(\x) \\
    &\textrm{subject to} & \sum_{a \in \A} x_{a,c} &\leq q_c & \text{ for all } \; c \in \C \\
    && \sum_{c \in \C} x_{a,c} &\leq 1 & \text{ for all } \; a \in \A \\
    && x_{a,c} &= 0 & \text{ for all } \; a \in \A, c \in \C \text{ with } a \notin \E_c \\
    && x_{a,c} &\geq 0 & \text{ for all } \; a \in \A, c \in \C.
\end{align*}

The decision variables $\mathbf{x} = (x_{a,c})_{a \in \A, c \in \C}$ represent the amount allocated to agent $a$ through category $c$. The three sets of constraints enforce \ax{\quota}, the unit demand of agents, and \ax{\eligibility}, respectively. The objective, $V(\x) := \sum_{a \in \A} \sum_{c \in \C} x_{a,c}$ is the total allocation of the agents; maximizing $V(\x)$ ensures Pareto efficiency. Note that the constraints of $(P_0)$ encode a bipartite $b$-matching polytope. 
$(P_0)$, however, does not incorporate the category's priorities, so its solutions may not satisfy \ax{\priority}. Adding constraints that enforce respect for priorities appears non-trivial. In particular, the set of valid allocations is not even closed under convex combinations, as demonstrated in \cref{fig:non-convex}.

\begin{figure}[t]
    \begin{minipage}{0.22\textwidth}
    \centering
    $\x$: \quad
    \begin{tabular}{c|c}
         $\alpha$ {\scriptsize (2)} & $\beta$ {\scriptsize (2)} \\ \hline
         $\boxed{a}$ & $a$ \\
         $\boxed{b}$ & $b$ \\
         $c$ & $\boxed{e}$ \\
         $d$ & $\boxed{f}$
    \end{tabular}
    \end{minipage}
    \quad
    \begin{minipage}{0.22\textwidth}
    \centering
    $\y$: \quad
    \begin{tabular}{c|c}
         $\alpha$ {\scriptsize (2)} & $\beta$ {\scriptsize (2)} \\ \hline
         $a$ & $\boxed{a}$ \\
         $b$ & $\boxed{b}$ \\
         $\boxed{c}$ & $e$ \\
         $\boxed{d}$ & $f$
    \end{tabular}
    \end{minipage}
    \hfill
    \begin{minipage}{0.5\textwidth}
        \caption{\small\it
            Consider the two integral allocations $\x$ and $\y$ depicted for the above allocation instance (taken from \cite{delacretaz2021processing}). Both $\x$ and $\y$ are valid; however, the fractional allocation $\z = \tfrac{1}{2} (\x + \y)$ does not respect priorities: in particular category $\alpha$ gives agent $d$ an allocation $z_{d,\alpha} = \tfrac{1}{2} > 0$, but agent $c \succeq_\alpha d$ is not fully allocated.
            \label{fig:non-convex} \\
        } 
    \end{minipage}
\vspace{-0.5cm}
\end{figure}

The critical observation is that one can perturb the coefficient of each $x_{a,c}$ in the objective to $1-\delta_{a,c}$ in such a way as to ensure that \emph{any optimal solution} to the perturbed LP satisfies all of the desiderata. To do so, we introduce the notion of a \emph{valid perturbation}.
\medskip

\begin{tcolorbox}
\begin{definition}[\bf Valid Perturbation]
\label{def:perturb}
    A perturbation profile $( \delta_{a,c} )$ is \emph{valid} if it satisfies the following three properties:
    \medskip
    \begin{description}[nosep]
        \item[Positivity:] $\qquad\delta_{a,c} > 0\text{ for all } a \in \A, c \in \C$.
        \medskip
        \item[Small Effect:] $\;\;\sum\limits_{a \in \A} \sum\limits_{c \in \C} \delta_{a,c} \leq \frac{1}{2}$.
        \item[Consistency:] $\;\;\;a \succeq_c a'$ if and only if $\delta_{a',c} \geq \delta_{a,c}$.
    \end{description}
\end{definition}
\end{tcolorbox}
\noindent Now, consider the modified objective 
\[
    V_\delta(\x) := \sum_{a \in \A} \sum_{c \in \C} x_{a,c} \cdot \big(1 - \delta_{a,c} \big) = V(\x) - \sum_{a \in \A} \sum_{c \in \C} \delta_{a,c} \cdot x_{a,c}.
\]

Let $(P_\delta)$ be the LP with the same constraint polytope as $(P_0)$, but with objective $V_\delta(\x)$. The following theorem shows that the solutions to any such perturbed LP give allocations satisfying all of our desiderata. 

\begin{theorem} 
\label{thm:delta_perturb}
    Let $\delta$ be any valid perturbation profile, and let $\x^*$ be a solution to $(P_\delta)$. Then, $\x^*$ is a valid allocation (i.e., it satisfies \emph{\ax{\quota}}, \emph{\ax{\eligibility}}, \emph{\ax{\priority}}, and \emph{\ax{\pareto}}). 
\end{theorem}

\begin{proof}
    The constraints immediately ensure that any feasible solution of $(P_\delta)$ satisfies \ax{\quota} and \ax{\eligibility}. To establish \ax{\priority}, let $\x$ be a feasible solution, $a,a'$ be agents and $c$ a category such that $a' \succ_c a$, $x_{a,c} = \varepsilon_1 > 0$ and $\sum_{c'} x_{a',c'} = 1 - \varepsilon_2 < 1$. Then, we can decrease $x_{a,c}$ and {increase} $x_{a',c}$ by $\min(\varepsilon_1, \varepsilon_2)$ without violating any constraints. Since $\delta$ is consistent, we have $\delta_{a,c} < \delta_{a',c}$, so the reassignment strictly increases the objective value. Thus, such an $\x$ is not optimal, and $\x^*$, being optimal, satisfies \ax{\priority}. 
    
    It remains to establish \ax{\pareto}. Note that for any optimal solution $\hat{\x}$ to $(P_0)$, we have
    \[
        V(\x^*) \geq V_\delta(\x^*) \geq V_\delta(\hat{\x}) = V(\hat{\x}) - \sum_{a \in \A} \sum_{c \in \C} \hat{x}_{a,c} \delta_{a,c} 
        \geq V(\hat{\x}) - \sum_{a \in \A} \sum_{c \in \C} \delta_{a,c} 
        \geq V(\hat{\x}) - \tfrac{1}{2}.
    \]
    Here, the first inequality follows since each $x^*_{a,c}, \delta_{a,c} \geq 0$. The second inequality follows since $\x^*$ is an optimal solution to $(P_\delta)$. The third inequality follows because the unit demand constraints ensure that each $\hat{x}_{a,c} \leq 1$. Finally, the fourth inequality follows since $\delta$ has small effect.
    
    Additionally, $\hat{\x}$ maximizes $V$ among all feasible solutions to $(P_0)$, which include $\x^*$. Therefore, $V(\hat{\x}) \geq  V(\x^*)$.
    Combining both inequalities, we find that 
    \begin{equation}
    \label{eq:fractionalbounds}
        V(\hat{\x}) \geq  V(\x^*) \geq V(\hat{\x}) - \tfrac{1}{2}.
    \end{equation}
    Observe that the constraint matrix of $(P_0)$ is \emph{totally unimodular}, as it encodes a $b$-matching polytope. Consequently, since all of the quotas $q_c$ are integral, every corner point of the constraint polytope is integral. As the sum of entries $(x_{a,c})$, $V(\x)$ is integral at corner points, and therefore at all maximizers $\x$ of $V$.
    In particular, since $\hat{\x}$ maximizes $V$, $V(\hat{\x})$ is integral. 
    If $\x^*$ is a corner point, then $V(\x^*)$ is also integral.
    However, integral solutions satisfying the bounds in~\cref{eq:fractionalbounds} require $V(\hat{\x}) = V(\x^*)$.
    
    If $\x^*$ is not a corner point, then
    we write $\x^* = \sum_i \lambda_i \x^{(i)}$ as a convex combination of corner points $\x^{(i)}$. Because $\x^*$ maximizes $V_{\delta}$, each of the $\x^{(i)}$ must also maximize $V_{\delta}$. By the argument from the previous paragraph, $V(\hat{\x}) = V(\x^{(i)})$ for all $i$. But then, the convex combination $\x^*$ must also have $V(\x^*) = V(\hat{\x})$. Thus, each maximizer $\x^*$ of $V_{\delta}$ (whether or not it is a corner point) is also a maximizer of $V$, and hence satisfies \ax{\pareto}.
\end{proof}

A surprising consequence of this result is that in any priority-respecting allocation problem, Pareto efficiency comes ``for free'' --- the total allocation size under a valid integral allocation remains the same, irrespective of the priority orderings! The following proposition asserts that this is in fact true more generally for \emph{any} valid allocation (integral or fractional).

\begin{proposition} \label{prop:pe_is_max}
    Let $V^*$ be the size  of the allocation returned by $(P_0)$ (i.e., satisfying \emph{\ax{\quota}}, \emph{\ax{\eligibility}}, and \emph{\ax{\pareto}}). Then, given any priority orders $( \succeq_c )_{c \in \C}$, any valid allocation $\x$ has $V(\x)=V^*$.
\end{proposition}

In particular, this proposition establishes that all Pareto efficient allocations are maximal.

\begin{proof}
    We will argue the contrapositive --- i.e., any $\x$ that does not maximize $V(\x)$ is not \ax{\pareto}. Consider the flow network representation of the allocation problem shown in \cref{fig:flow_net}. The nodes on the left side correspond to the agents $a\in\A$, and the nodes on the right to categories $c \in \C$. Edges are drawn between each eligible agent-category pair.
    Finally, given an allocation $\x$, for every category $c$ that has an eligible agent $a\in\E_c$ who is not fully allocated in $\x$, we color all its eligible agents (i.e., all $a'\in\E_c$) red, whether or not they are fully allocated.

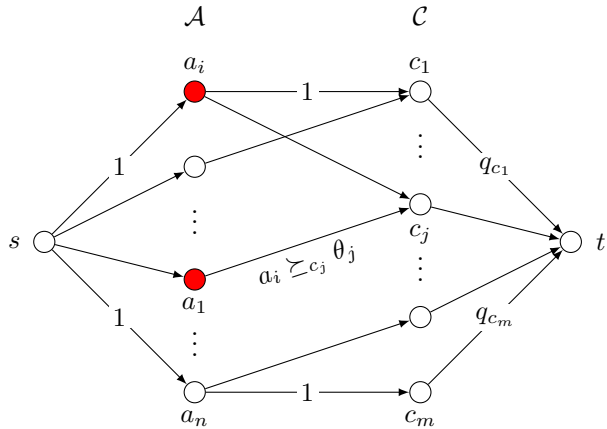
\begin{figure}[H]
    \centering
    \begin{tikzpicture}
        \node[draw, circle, inner sep=0pt, minimum size = 8pt] (s) at (0,2) {};
        \node[draw, circle, inner sep=0pt, minimum size = 8pt] (a1) at (2,0) {};
        \node[draw, circle, inner sep=0pt, minimum size = 8pt, fill=red] (a2) at (2,1.5) {};
        \node[draw, circle, inner sep=0pt, minimum size = 8pt] (a3) at (2,3) {};
        \node[draw, circle, inner sep=0pt, minimum size = 8pt, fill=red] (a4) at (2,4) {};
        \node[draw, circle, inner sep=0pt, minimum size = 8pt] (c1) at (5,0) {};
        \node[draw, circle, inner sep=0pt, minimum size = 8pt] (c2) at (5,1) {};
        \node[draw, circle, inner sep=0pt, minimum size = 8pt] (c3) at (5,2.5) {};
        \node[draw, circle, inner sep=0pt, minimum size = 8pt] (c4) at (5,4) {};
        \node[draw, circle, inner sep=0pt, minimum size = 8pt] (t) at (7,2) {};
        
        \draw[-latex] (s) -- (a1); 
        \draw[-latex] (s) -- (a2);
        \draw[-latex] (s) -- (a3);
        \draw[-latex] (s) -- (a4);
        
        \draw[-latex] (c1) -- (t); 
        \draw[-latex] (c2) -- (t);
        \draw[-latex] (c3) -- (t);
        \draw[-latex] (c4) -- (t);
        
        \draw[-latex] (a1) -- (c1);
        \draw[-latex] (a4) -- (c4);
        \draw[-latex] (a3) -- (c4);
        \draw[-latex] (a2) -- (c3);
        \draw[-latex] (a1) -- (c2);
        \draw[-latex] (a4) -- (c3);
        
        \node at (2,5) {$\mathcal{A}$};
        \node at (5,5) {$\mathcal{C}$};
        \node at (-0.4,2) {$s$};
        \node at (7.4,2) {$t$};
        
        \node at (2,0.75) {$\vdots$};
        \node at (2,2.375) {$\vdots$};
        \node at (5,1.75) {$\vdots$};
        \node at (5,3.375) {$\vdots$};
        
        \node[fill=white, inner sep=0pt, minimum size=8pt] at (1,1) {$1$};
        \node[fill=white, inner sep=0pt, minimum size=8pt] at (1,3) {$1$};
        \node[fill=white, inner sep=0pt, minimum size=10pt] at (6,3) {$q_{c_1}$};
        \node[fill=white, inner sep=0pt, minimum size=10pt] at (6,1) {$q_{c_m}$};
        \node[fill=white, inner sep=0pt, minimum size=8pt] at (3.5,0) {$1$};
        \node[fill=white, inner sep=0pt, minimum size=8pt] at (3.5,4) {$1$};
        
        \node[rotate=18] at (3.5,1.7) {$a_i \succeq_{c_j} \theta_j$};
        
        \node at (2,4.35) {$a_i$};
        \node at (2,1.15) {$a_1$};
        \node at (2,-0.35) {$a_n$};
        \node at (5,2.1) {$c_j$};
        \node at (5,4.35) {$c_1$};
        \node at (5,-0.35) {$c_m$};
    \end{tikzpicture}
    \caption{\small\it A flow network representation of an allocation instance. The source node $s$ has a unit-capacity edge to each agent node. Each category node has an edge to the sink node $t$ with capacity equal to that category's quota. There are unit-capacity edges from each agent node to the nodes of categories in which the agent is eligible.}
    \label{fig:flow_net}
    \end{figure}

    If $\x$ is not a maximal allocation, then there is an augmenting path $P = (s,a_1,c_1,\ldots,a_k,c_k,t)$ in this flow network. We record the following observations.
    \begin{enumerate}
        \item $a_1$ is red: The in-weight of each agent node is its allocation. Augmenting along $P$ will increase the in-weight of its first agent node, so this agent node must not have been fully allocated. 
        \item $c_k$ has not exhausted its quota: The out-weight of each category node is its allocated quota. Augmenting along $P$ will increase the out-weight of its last category node, so this category must not have exhausted its quota.
        \item Given any red agent $a$, there is a path of the form $s \to a_0 \to c_0 \to a$ in the residual graph for $\x$, where $a_0$ is a highest-priority agent in $c_0$ that is not fully allocated: this follows from the definition of red agent nodes.
    \end{enumerate}

    Let $a_i$ be the \emph{last} red node in $P$ (there must be such a node by Observation 1), and consider the alternate augmenting path $P' = (s,a_0,c_0,a_i,c_i,\ldots, a_k,c_k,t)$ using the ``shortcut'' from Observation 3. Augmenting along $P'$ will strictly increase the allocation to $a_0$ and conserves the allocations of $a_i,\ldots,a_k$. Let $\y$ be the allocation after this augmentation. By the construction of the flow network, $\y$ still satisfies \ax{\eligibility} and \ax{\quota}. Moreover, every agent $a \succeq_{c_0} a_0$ is fully allocated, and every agent {$a_i, \ldots, a_k$} maintains its allocation in $\y$, so $\y$ also satisfies \ax{\priority}. Thus, $\y$ is a Pareto improvement to $\x$, meaning $\x$ did not satisfy \ax{\pareto}. 
\end{proof}

The fact that there is at least one valid allocation with size $V^*$ was established by~\citet{aziz2021efficient} based on the properties of their \textsc{Reverse Rejection} algorithm. \cref{thm:delta_perturb} gives a simple way to see why this holds, and~\cref{prop:pe_is_max} shows it to be true for all valid allocations. Moreover, \cref{thm:delta_perturb} provides a much more efficient algorithm for selecting a valid allocation: compared to \textsc{Reverse Rejection}, which requires one to solve $|\A|$ separate $b$-matching problems, our approach only requires solving a single weighted $b$-matching problem, which can be efficiently solved, for instance using the Hungarian algorithm~\citep{ramshaw2012minimum}.

\begin{corollary} \label{cor:runtime}
    A valid allocation can be found in $O(|\C|\,|\A|\,q + q^2\log q)$ time.
\end{corollary}

\subsection{The Subtle Geometry of Priority-Respecting Allocations}
\label{ssec:characterization}
To conclude this section, we discuss three issues related to our algorithm for locating valid allocations. First, we introduce an additional property that allows us to completely characterize the set of valid integer allocations. Next, we consider the geometry of the set of valid allocations. Finally, we consider whether an analogous LP perturbation can be used for finding stable matchings.

\medskip
\noindent\textbf{Characterizing all Valid Integral Allocations}\\
By Theorem~\ref{thm:delta_perturb}, we know that solving $(P_\delta)$ with any valid $\delta$  locates a valid integral allocation. 
A follow-up question is whether \emph{all} valid integral allocations are solutions of $(P_\delta)$ for some choice of $\delta$. This turns out not to be the case: for example, consider Allocations $3$ and $4$ in~\cref{fig:axioms}. While both are valid, and the former can be realized as a solution to a perturbed LP, for the latter allocation, under any valid perturbation, swapping from $x_{a,\gamma}=x_{b,\beta}=1$ to $x_{b,\gamma}=x_{a,\beta}=1$ (as in Allocation 3 in~\cref{fig:axioms}) leads to an increase in $V_\delta$.

Fortunately, the problem illustrated by this instance is the only obstacle to realizability, as we show below. To formalize this, we introduce an additional property that we call category stability. 

\medskip
\begin{description}
    \item[{[\stability]} Category Stability:] 
    No group of categories can organize an agreeable trade through which at least one category transfers allocation to a higher-priority agent. Formally, there do not exist $j \geq 2$ categories $c_0, c_1, \hdots, c_j = c_0$ and agents $a_0, a_1, \hdots, a_j = a_0$ such that for all $0 \leq i < j$, $x_{a_i,c_i} > 0$ and $a_{i+1} \succeq_{c_i} a_i$, and at least one of the priority relations above is strict.
\end{description}
Note that category stability is not an added restriction on \emph{agents} selected via valid allocations --- in particular, given a valid allocation $\x$ that violates \ax{\stability}, we can modify it to get another valid allocation $\y$ that satisfies \ax{\stability} and \emph{allocates to each agent to the same extent} (i.e., $\sum_cx_{a,c}=\sum_c y_{a,c}$ for all $a$). In other words, \ax{\stability} only discriminates among valid allocations which are equivalent in terms of the set of allocated agents. We are ready to state our main equivalence theorem.

\begin{theorem} \label{thm:realizability}
    Let $\x$ be a valid integral allocation. Then $\x$ is a solution to $(P_\delta)$ for some valid $\delta$ \emph{if and only if} $\x$ satisfies \emph{\ax{\stability}}.
\end{theorem}

The main tool we use to show the reverse implication of~\cref{thm:realizability} is an alternate characterization of the valid allocations that additionally satisfy \ax{\stability} as those realizable through \emph{serial dictatorship}. Let $\Sigma$ be the collection of all multi-set orderings of $\big\{ c^{q_c} \big\}_{c \in \C}$ (i.e., the set of all sequences of length $q$ wherein each category $c\in\C$ appears $q_c$ times). We refer to $\Sigma$ as the set of {choice orders} for our system. For a given choice order $\sigma \in \Sigma$, we define the \emph{serial dictatorship allocation} $\x_\sigma$ to be the (integral) allocation obtained by cycling through categories in the order given by $\sigma$, and allocating to the highest-priority unallocated agent in the chosen category. This process is formalized in~\cref{alg:serial_dict}. For ease of presentation, we ignore ties in~\cref{alg:serial_dict}. This assumption corresponds to each category having a total ordering over eligible agents; in case there are multiple unallocated agents in the same highest-priority tier, we can use any fixed tie-breaking rule (alternately, any fixed extension of the total preorder $\succeq_c$).
\begin{algorithm}[H]
    \caption{Serial Dictatorship Allocation}
    \textbf{Input}: Choice order $\sigma \in \Sigma$ \hfill
    \begin{algorithmic}[1] \label{alg:serial_dict}
        \FOR{each $\sigma_i=c$ in $\sigma$ in order}
        \IF{$c$ has remaining quota and an eligible unallocated agent}
            \STATE $c$ allocates to its highest-priority unallocated agent. 
        \ENDIF
        \ENDFOR
    \end{algorithmic}
\end{algorithm}

Serial dictatorship allocations $x_{\sigma}$ generalize the sequential reserve allocations of~\citet{pathak2021fair}. It is straightforward to see that they (by definition) satisfy \ax{\quota}, \ax{\eligibility} and \ax{\priority}. Note, however, that $\x_\sigma$ may not be Pareto efficient (for example, consider allocation 2 in~\cref{fig:axioms} --- it can be realized as a serial dictatorship allocation $\x_\sigma$ with $\sigma = (\beta,\gamma,\alpha)$.) The following lemma fully characterizes the allocations obtained via serial dictatorship and generalizes a main result of~\citet{pathak2021fair}.

\begin{lemma} \label{lem:serialdictator}
    For all $\sigma \in \Sigma$, the serial dictatorship allocation $\x_\sigma$ satisfies \emph{\ax{\quota}}, \emph{\ax{\eligibility}}, \emph{\ax{\priority}}, and \emph{\ax{\stability}}. Conversely, every valid integral allocation (i.e., obeying \emph{\ax{\quota}}, \emph{\ax{\eligibility}}, \emph{\ax{\priority}}, and \emph{\ax{\pareto}}) that additionally satisfies \emph{\ax{\stability}} corresponds to a serial dictatorship allocation  $\x_\sigma$ under some choice order $\sigma \in \Sigma$.
\end{lemma}

\begin{proof}
    For the first claim, it is immediate from the definition of serial dictatorship that $\x_\sigma$ satisfies \ax{\quota}  (since $\sigma$ contains $q_c$ copies of $c$), \ax{\eligibility} (since each category $c$ only allocates to eligible agents), and \ax{\priority} (since a category always allocates to a highest-priority unallocated agent). To see that $\x$ is stable, for any subset $S$ of allocated agents consider the first time that agent $a\in S$ is allocated by a category $c$. By definition, $c$ selects a highest-priority unallocated agent, so $a \succeq_c s$ for all $s \in S$. Thus, $S$ cannot form an unstable cycle.  
    
    To show the second claim (that every valid integral allocation satisfying \ax{\stability} can be generated via a serial dictatorship allocation), we perform an induction on $q$. The base case $q=1$ is trivial: if $c$ is the category with $q_c=1$, then any valid allocation that also satisfies \ax{\stability} must give this unit to a highest-priority eligible agent in $c$, if one exists. 
    
    Suppose that the claim holds for all instances with $q = k-1$, and consider an instance with quota $q=k$. We first show that in any valid and \ax{\stability} allocation $\x$ (with $V(\x) > 0$), a highest-priority agent in some category is allocated from that category. Suppose that this were not the case, and consider an agent $a$ who is allocated from category $c$. By assumption, there is some highest-priority agent $a'$ who is not allocated from $c$. If $a'$ is unallocated, then $\x$ would violate \ax{\priority}. Hence, $a'$ must be allocated in some other category $c'$. By assumption, $a'$ does not have highest priority in $c'$, meaning that the highest-priority agent $a''$ of $c'$ is not allocated in $c'$. Continuing this reasoning, we will (by finiteness) eventually revisit an agent and discover an unstable cycle, contradicting that $\x$ satisfies \ax{\stability}. 
    
    Now, let $c^*$ be a category allocating to its highest-priority agent, and $a^*$ the highest-priority agent in $c^*$. We can realize this allocation by having $c^*$ be the first category in the ordering $\sigma$. What remains is an allocation problem for agents $\A \setminus \{a^*\}$ to categories $\C$, where the quota of $c^*$ has been reduced by 1. Let $\y$ be the restriction of $\x$ to this problem. It is immediate that $\y$ is a valid and \ax{\stability} allocation. By our inductive hypothesis, $\y$ can be realized as a serial dictatorship allocation $\y_{\sigma'}$ in this sub-problem. Then, $\x_{(c^*,\sigma')}$ realizes $\x$.
\end{proof}

Using this lemma, we can complete the proof of~\cref{thm:realizability}. 

\begin{proof}[Proof of \cref{thm:realizability}]

    For the forward direction, we argue the contrapositive. Suppose that $\x$ is feasible for $(P_\delta)$ for some valid $\delta$. Suppose that $\x$ violates \ax{\stability}, so there are $a_0, a_1, \ldots, a_j = a_0 \in \A$ and  $c_0, c_1, \ldots, c_j = c_0 \in \C$ for which $x_{a_i,c_i} = 1$, $x_{a_{i+1},c_i} = 0$, and $a_{i+1} \succ_{c_i} a_i$ for each $0 \leq i < j$. We construct an alternate solution $\x'$ with $x'_{a_i,c_i} = 0$ and $x_{a_{i+1},c_i} = 1$ for each $0 \leq i < j$ and all other variables the same as $\x$. Note that $\x'$ is also feasible since $a' \succ_c a \implies a' \in \E_c$, and all categories and agents have the same total allocation. Since $\delta$ is consistent, we have $\delta_{a_{i+1},c_i} < \delta_{a_i,c_i}$ for each $0 \leq i < j$, so the reassignment strictly increases the objective value. Thus, $\x$ is not optimal, so it is not a solution to $(P_\delta)$.
    
    For the reverse direction, we must construct an assignment of perturbations $\delta$ that realize the allocation $\x$ as a solution. It will be convenient to argue using \emph{positive} perturbations (i.e., a bonus rather than a penalty). That is, for every $a\in\A,c\in\C$, we set the coefficient of $x_{a,c}$ in the objective as $1+\rho_{a,c}$, such that $\rho_{a,c}\in[0,\rho_{\max}]$ for all eligible $(a,c)$, and $\rho_{a,c}\geq \rho_{a',c}$ for all $a\succeq_c a'$. To convert the $\rho_{a,c}$ to valid perturbations $\delta_{a,c}$ (\cref{def:perturb}), we can simply re-scale them by $\tfrac{1}{1+\rho_{\max}}$ to get $\delta_{a,c} = \frac{\rho_{\max}-\rho_{a,c}}{1+\rho_{\max}}$.  Then, it is easy to check that these perturbations satisfy Positivity and Consistency. Also, by choosing $\rho_{\max}=\frac{1}{2|\C|\,|\A|}$, we ensure that $\sum_{a,c} \delta_{a,c} \leq |\C| |\A| \cdot \rho_{\max} /(1+\rho_{\max}) \leq 1/2$; thus, the $\delta_{a,c}$ constitute a valid perturbation.
    
    Let $v := V(\x)$. By~\cref{lem:serialdictator}, $\x = \x_\sigma$ for some ordering $\sigma = (\sigma_1, \ldots, \sigma_q) \in \Sigma$. 
    We may also, without loss of generality, assume that the first $v$ entries of $\sigma$ result in the allocation of an agent: note that any entry $\sigma_i$ corresponding to a depleted category can be moved to the end of the ordering without affecting the agents available to any later entry.
    
    Now, we set the perturbations as follows:
    \begin{enumerate}
    \item Let $a$ be the top-ranked agent in the category $\sigma_1$. We set $\rho_{a,\sigma_1} = \rho_{\max}$.
    \item In stage $i$, let $r \leq i$ be the lowest rank of an unallocated agent in category $\sigma_i$. 
    Let $r' < r$ be the rank of the agent most recently allocated in $\sigma_i$, with $r'=0$ if no agent has yet been allocated through $\sigma_i$.
    For $j=r'+1, r'+2, \ldots, r$, let $a_j$ be the agent with rank $j$ in $\sigma_i$, and define $A_i = \{ a_{r'+1}, a_{r'+2}, \ldots, a_r\}$.
    We set
    $\rho_{a_j, \sigma_i} = \rho_{\max}/(|\A|+1)^{i-1} + (r-j) \cdot \varepsilon$, 
    for some $\varepsilon \ll \rho_{\max}/(|\A|+1)^{|\A|}$.
    \end{enumerate}
    
    The main invariant maintained by the above construction is that at any stage $i$, the smallest perturbation $\rho_{a,c}$ for $c=\sigma_i$ and any $a \in A_i$ is greater than the \emph{sum of all perturbations} of $(a,c)$ pairs set in rounds $i' > i$. As a result, the optimal matching among pairs $(a,c)$ considered in rounds $i$ and greater must include at least one pair  $(a_j,\sigma_i)$ for some $a_j \in A_i$. Moreover, since the agents $a_{r'+1}, a_{r'+2}, \ldots, a_{r-1}$ were allocated in rounds prior to $i$, any optimal matching with respect to the $\rho_{a,c}$ must have $x_{a_r,\sigma_i}=1$. This exactly corresponds to the outcome $x_{\sigma}$ realized via Serial Dictatorship with order $\sigma$. Thus, $x_\sigma$ is realized as a solution to $(P_\delta)$.  
\end{proof}

\medskip
\noindent\textbf{Fractional Valid Allocations}\\
Our perturbed LP procedure gives a way to locate all valid and category stable \emph{integral} allocations, as these are corner points of our $b$-matching constraint polytope; however, this does not imply anything about the set of valid \emph{fractional} allocations. \cref{fig:non-convex} demonstrates that this set need not be convex; here, the convex combination of two integral valid allocations is not valid. Thus, there could exist valid fractional allocations outside the convex hull of valid integral allocations. The following proposition rules out this possibility.

\begin{proposition}
\label{prop:fractionalvalid}
    Suppose that $\x$ is a valid fractional allocation. Then, we can represent $\x$ as a convex combination of valid integer allocations. 
\end{proposition}

The following lemma will be useful in our proof of \cref{prop:fractionalvalid}.

\begin{lemma} 
\label{lem:serv_int}
Consider any valid allocation $\x$ with allocated agents $\A_x=\{a\in\A \colon \sum_{c}x_{a,c}>0\}$. Then, for any agent $a^* \in \A_x$ who is partially allocated (i.e., $0 < \sum_{c} x_{a^*,c} < 1$), there exists a valid allocation $\y$ with $\A_y\subseteq \A_x $ and in which $a^*$ is fully allocated (i.e., $\sum_{c} y_{a^*,c} = 1$). 
\end{lemma}

\begin{proof}
Let $c^* \in \C$ be any category with $0 < x_{a^*,c^*} < 1$. We argue that there is a way to modify $\x$ which maintains validity, strictly increases $x_{a^*,c^*}$, and strictly decreases the number of non-integral allocation variables. Since the number of eligible category-agent pairs (and therefore, the number of non-integral allocations) is finite, we can repeatedly apply this modification until 
$a^*$ is fully allocated.
    
To describe the modification, we first construct an undirected graph as follows.
    \begin{itemize}
        \item[---] The nodes of the graph will correspond to $(a,c)$ pairs with $0 < x_{a,c} < 1$. 
        \item[---] We color an agent $a$, and all its associated nodes, red if it is not fully allocated (i.e. $\sum_{c' \in \C} x_{a,c'} < 1)$, otherwise white.
        \item[---] We add an edge between any two nodes that share a category.
        \item[---] We add an edge between any two \emph{white} nodes that share an agent.
    \end{itemize}
    Note that the third bullet implies that the connected components of this graph describe a partition of the categories. We argue that the red node $(a^*,c^*)$ is in the same connected component as another red node. For sake of contradiction, suppose not. Note that the total quota of all categories associated with this component is an integer. In addition, the total allocation to all of the agents associated with this component is not an integer: the white agents each have allocation 1, and the singular red agent has a non-integral allocation. However, all of the quotas must be exhausted by the allocation. If not, a path from $(a^*,c^*)$ to a node $(a,c)$ where $c$ has not exhausted its capacity describes a way to adjust the allocation to increase the total allocation to $a^*$ and leave all other agents' total allocations unchanged, violating Pareto efficiency. However, this is a contradiction: the total quota of these categories cannot be both integral and non-integral.

    Suppose that $(\hat{a},\hat{c})$ is another red node in $(a^*,c^*)$'s connected component. By definition, there is a path between these two nodes. Moreover, the structure of the graph allows us to assume (without loss of generality) that the edges in this graph alternate between connecting nodes that share an agent and nodes that share a category. Since red nodes are only connected to nodes with which they share a category, this path has an odd length. We modify the allocation by following the path from $(\hat{a},\hat{c})$ to $(a^*,c^*)$. First, we subtract $\varepsilon > 0$ from $x_{\hat{a},\hat{c}}$. Then, we add $\varepsilon$ to the variable corresponding to the next node on the path, repeating this process until we add $\varepsilon$ to $(a^*,c^*)$: the first bullet point allows us to choose $\varepsilon$ such that one of these modifications results in a variable assuming value in $\{0,1\}$. 
    
    To finish the proof, we must argue that this modification results in another valid allocation, which we denote by $\x'$. First, note that the modification did not change the total allocation of any category; it only transferred quota from one agent to another. Thus, $\x'$ satisfies \ax{\quota}. In addition, we conclude by \cref{prop:pe_is_max} that $\x'$ satisfies \ax{\pareto}: it is also a maximal allocation. Next, note that the modification does not transfer any quota to an $(a,c)$ node with $x_{a,c} = 0$, so $\x'$ satisfies \ax{\eligibility}. Finally, note that the only agent whose total allocation can decrease from the modification is $\hat{a}$, who is red. Therefore, $\x'$ maintains \ax{\priority}.  
\end{proof}

We are now ready to prove \cref{prop:fractionalvalid}.

\begin{proof}[Proof of \cref{prop:fractionalvalid}]
    We argue this claim in two stages. First, we argue that $\x$ can be represented as a convex combination of valid allocations in which each agent has an integer total allocation. This follows from \cref{lem:serv_int}. In this proof, we obtained an alternate allocation $\x'$ from $\x$ by perturbing nodes along a path by $\varepsilon$. Similarly, add $\varepsilon' > 0$ to $x_{\hat{a},\hat{c}}$, subtract $\varepsilon'$ from the next node, repeating until we subtract $\varepsilon'$ from $(a^*,c^*)$ to obtain an alternate valid allocation $\x''$: the first bullet point allows us to choose $\varepsilon'$ such that one of these modifications results in a variable assuming value in $\{0,1\}$. But then we can express $\x$ as the convex combination
    \[
        \x = \tfrac{\varepsilon'}{\varepsilon+\varepsilon'} \cdot \x' + \tfrac{\varepsilon}{\varepsilon+\varepsilon'} \cdot \x''.
    \]
    We can repeat this process with $\x'$ and $\x''$, just as in the proof of \cref{lem:serv_int}. Since each step strictly decreases the number of non-integral variables, eventually, we will be left with a convex combination of valid allocations $\{\y^{(1)}, \hdots, \y^{(\ell)} \}$ in which each agent has an integer total allocation. (This is the termination condition of the procedure from \cref{lem:serv_int}.) To conclude, we must further represent each $\y^{(i)}$ as a convex combination of valid integral allocations (i.e., allocations in which each allocated agent receives an entire unit from exactly one category). This is an application of the Birkhoff-von Neumann theorem: since each agent is fully allocated, we can interpret $\y^{(i)}$ as fractional matchings between the agents and categories in the subgraph of edges $(a,c)$ with $\y^{(i)}_{a,c} > 0$. Validity is preserved since these integer matchings preserve the total allocation to each agent and category.
\end{proof}

The result, however, provides some insight into the geometry of the set of valid allocations --- it consists of a union of convex sets, each with integer corner points. 
The valid allocations in the example from \cref{fig:non-convex} form two non-coplanar triangles with a common edge. It is an interesting open direction to further characterize the sets of valid allocations that may arise from priority-respecting allocation instances. For example, are these sets necessarily connected?

\medskip
\noindent\textbf{LP Perturbations and Stable Matching}

As noted in \cref{sec:intro}, a closely related problem to priority-respecting allocation is stable matching.
Both seek bipartite matchings that conform to a set of preferential constraints; moreover, the existence of solutions in both problems is implied by Scarf's Lemma (see~\cref{sec:scarf}). Central to this discussion is the observation that unlike for priority-respecting allocation, one cannot realize stable matchings as the solutions of a perturbed $b$-matching polytope under a particular class of perturbations. This shows that while priority-respecting allocations and stable matchings appear syntactically similar, they have very different algorithmic properties.

%% file: s4a_hardness.tex
\section{The Complexity of Selecting Valid Allocations} 
\label{sec:examples}
  
In this section, we consider three possible extensions of the basic problem of selecting valid allocations. In each extension, we consider the problem of selecting from among valid allocations subject to a particular class of external objectives. 
Surprisingly, in each case, we show that the valid-allocation selection problem straddles the line of computational efficiency; while one given selection rule admits an efficient algorithm, a closely related selection rule is computationally hard.

\subsection{Including/Excluding Agents from Valid Allocations}
\label{sec:serviceable}

To formalize our first set of valid-allocation selection criteria, we first define two types of agents

\begin{definition}[\bf Unanimous/Serviceable Agents] 
Given an instance $\mathcal{I} = \big( \A, \C, (q_c), (\E_c), (\succeq_c) \big)$ and an agent $a \in \A$,
\begin{itemize}[leftmargin=*]
\item[---] Agent $a$ is \emph{unanimous} in $\mathcal{I}$ if it is fully allocated under \emph{every} valid allocation $\x$ 
(i.e., $\sum_cx_{a,c}=1$).  
\item[---] Agent $a$ is \emph{serviceable} in $\mathcal{I}$ if there is some valid allocation $\x$ in which $a$ is allocated (i.e., $\sum_c x_{a,c} >0$). 
\end{itemize}
\end{definition}

As an example, consider the instance in~\cref{fig:axioms}; here, one can check that agents $\{a,b,c\}$ are unanimous (and therefore serviceable), while agent $d$ is \emph{not} serviceable.
Note that though we define serviceability in terms of non-zero allocation, as a consequence of~\cref{prop:fractionalvalid}, we have that any agent who can be partially allocated via a fractional valid allocation can also be fully allocated via an integral valid allocation. 
Hence, we can equivalently define an agent to be serviceable if it is allocated in some integral allocation.

We now show that while there is a polynomial-time algorithm to determine whether an agent is unanimous, determining whether an agent is serviceable is \textsf{NP}-hard.
For the first claim, we establish an equivalent characterization of unanimous agents in terms of a \emph{restricted} allocation instance.
\begin{definition}[\bf Restriction] \label{def:restrict}
    Given an allocation instance $\mathcal{I} = \big( \A, \C, (q_c), (\E_c), (\succeq_c) \big)$ and an agent $a \in \A$, the \emph{$a$-restriction} of c, denoted by $\mathcal{I}_{\setminus a}$, is another allocation instance with the same $\A$, $\C$, and $(q_c)$. Its eligible sets $(\E'_c)$ are given by
    \[
        \E'_c = \E_c \setminus \Big( \{ a \} \cup \{ a' \in \A \::\: a \succ_c a' \} \Big),     
    \]
    and its priorities $(\succeq'_c)$ are the induced relations of $(\succeq_c)$ on $(\E'_c)$. 

    Given a subset $A \subseteq \A$, the \emph{$A$-restriction} of $\mathcal{I}$, denoted by $\mathcal{I}_{\setminus A}$, is defined similarly, where the eligible sets are the intersections of the eligible sets of the $a$-restrictions for all $a \in A$. 
\end{definition}

Intuitively, one can think of the $a$-restriction as cutting each of the priority lists at agent $a$. Alternatively, one can view the $a$-restriction as the instance that would result if we committed to never allocating to $a$ (and therefore, due to the priority constraints, never allocating from a category $c$ to anyone ranked below $a$ in $c$). 

\begin{proposition}
\label{prop:unanimous}
    Let $V^*$ be the value of $(P_0)$ on instance $\mathcal{I}$. For a given agent $a \in \A$, let $V^*_{\setminus a}$ be the value of $(P_0)$ on $\mathcal{I}_{\setminus a}$. Then, $a$ is unanimous if and only if $V^* > V^*_{\setminus a}$.  
\end{proposition}

\begin{proof}
    We argue the forward direction by its contrapositive. Suppose that $V^* = V^*_{\setminus a}$, and let $\x$ be a solution to $(P_0)$ for the restricted instance $\mathcal{I}_{\setminus a}$. By~\cref{prop:pe_is_max}, there must be another solution $\y$ that additionally respects priorities (i.e., is valid). Note that $a$ is not eligible in any category in $\mathcal{I}_{\setminus a}$, so $y_{a,c}=0$ for every $c\in\C$. 
    However, $\y$ is also a valid allocation for the original instance $\mathcal{I}$: eligibility in $\mathcal{I}_{\setminus a}$ implies eligibility in $\mathcal{I}$; quota constraints are the same in $\mathcal{I}$ and $\mathcal{I}_{\setminus a}$; priorities are respected since the definition of restriction ensures that any eligible agent in $\mathcal{I}$ who is not fully allocated in $\y$ must be ranked below fully allocated agents in $\mathcal{I}_{\setminus a}$, and hence in $\mathcal{I}$; finally, $\y$ returns a matching of maximum size in $\mathcal{I}$, and so is Pareto efficient in $\mathcal{I}$. Thus we have located a valid allocation that does not include $a$, and hence $a$ is not unanimous.
    
    We also argue the reverse direction by its contrapositive. Suppose $a$ is not unanimous --- then, there is a valid allocation $\x$ in which $a$ is not allocated. By definition, this allocation has value $V^*$. Since $\x$ satisfies \ax{\priority}, no category can allocate to an agent with lower priority than $a$. Thus, $\x$ is feasible for $\mathcal{I}_{\setminus a}$, so $V^* = V^*_{\setminus a}$.
\end{proof}
 
As immediate corollaries, we can derive two sufficient conditions for an agent to be unanimous.
\begin{corollary}
     Let $V^*$ be the value of $(P_0)$. Then an agent $a$ is unanimous if either
     \begin{itemize}[nosep]
     \item[---] the union of all eligible agents in the $a$-restriction has cardinality less than $V^*$, or
    \item[---] there is some category $c$ with $a\in \E_c$ such that the $a$-restriction of $c$ has size less than $q_c$.
    \end{itemize}
\end{corollary}
In other words, an agent must be allocated if they are either in the top $q_c$ agents in any category $c$ or alternatively if the total (over categories) number of higher-ranked agents is less than the total number of items available. On the other hand, the problem of deciding whether an agent is \emph{serviceable} or not turns out to be \textsf{NP}-hard. 
\begin{proposition} \label{prop:servicible_hard}
    Given an instance $\mathcal{I}$, deciding whether an agent $a \in \A$ is serviceable is \textsf{NP}-hard.
\end{proposition}

\begin{proof}
We show this via a reduction from the \textsc{Exact Cover By 3-Sets}, or \textsc{X3C}, problem~\citep{karp1972reducibility}, which is defined as follows.
\begin{definition}[\textsc{X3C}]
    Given a ground set $E$ of $3n$ elements and a collection of $m$ subsets $\mathcal{S} = \{ S_1, \hdots, S_m \}$, with each $|S_i| = 3$, the \textsc{X3C} problem asks whether there are $n$ subsets $S_{i_1}, \hdots, S_{i_n}$ whose union is $E$.  
\end{definition}

We consider the following reduction from \textsc{X3C}, which is visualized in \cref{fig:x3c_reduction}. 
    \begin{itemize}
        \item[---] $\A$ consists of the following $5m-n+1$ agents: 
        \begin{itemize}
            \item $3n$ agents representing the ground set elements $e \in E$
            \item $m$ agents $s_1, \ldots, s_m$ representing the subsets $S_i \in \mathcal{S}$
            \item $4(m-n)$ \emph{filler agents}, labeled $f_1, \hdots, f_{4(m-n)}$
            \item the distinguished agent $a$ 
        \end{itemize}
        \item[---] $\C$ consists of $|\C|=m+1$ categories: a \emph{set category} $\alpha_i$ for each $S_i \in \mathcal{S}$ and a category $\beta$.
        \item[---] Each set category has quota 4, and $\beta$ has quota 1.
        \item[---] Each set category $\alpha_i$ has $4(m-n+1)$ eligible agents: the $4(m-n)$ filler agents, who have priority over agent $s_i$, who has priority over the 3 element agents in $S_i$.
        \item[---] The category $\beta$ has $3n+1$ eligible agents: the $3n$ element agents, who have priority over agent $a$. 
    \end{itemize}
    
    \begin{figure}[H]
        \centering
        \begin{minipage}{0.4\textwidth}
        \begin{align*}
            E &= \{ e_1,e_2,e_3,e_4,e_5,e_6 \} \\
            \mathcal{S} &= \Big\{ \{ e_1, e_3, e_6 \} , \\
            &\hspace{22pt} \{ e_1, e_4, e_5 \}, \\
            &\hspace{22pt} \{ e_2, e_4, e_5 \} \Big\}
        \end{align*}
        \end{minipage}
        $\longrightarrow$
        \begin{minipage}{0.4\textwidth}
        \begin{center}
        \begin{tabular}{c|c|c|c}
             $\alpha_1$ {\scriptsize (4)} & $\alpha_2$ {\scriptsize (4)} & $\alpha_3$ {\scriptsize (4)} & $\beta$ {\scriptsize (1)} \\ \hline
             $f_1$ & \cellcolor{red!15} $f_1$ & $f_1$ & $e_1$ \\
             $f_2$ & \cellcolor{red!15} $f_2$ & $f_2$ & $e_2$ \\
             $f_3$ & \cellcolor{red!15} $f_3$ & $f_3$ & $e_3$ \\
             $f_4$ & \cellcolor{red!15} $f_4$ & $f_4$ & $e_4$ \\
             \cellcolor{red!15} $s_1$ & $s_2$ & \cellcolor{red!15} $s_3$ & $e_5$ \\
             \cellcolor{red!15} $e_1$ & $e_1$ & \cellcolor{red!15} $e_2$ & $e_6$ \\
             \cellcolor{red!15} $e_3$ & $e_4$ & \cellcolor{red!15} $e_4$ & \cellcolor{red!15} $a$ \\
             \cellcolor{red!15} $e_6$ & $e_5$ & \cellcolor{red!15} $e_5$ &
        \end{tabular}
        \end{center}
        \end{minipage}
        \caption{An example reduction from an \textsc{X3C} instance (with $n=2$, $m=3$) to a reserve allocation instance. This is a ``yes'' instance of \textsc{X3C}: the first and third sets form a partition of $E$. Accordingly, the reserve allocation instance on the right admits a valid allocation, visualized in red, that gives to $a$.}
        \label{fig:x3c_reduction}
    \end{figure}

    This reserve allocation instance has size which is polynomial in $m$ and $n$, and it can be constructed in polynomial time. It remains to argue the correctness of the reduction. First, suppose that we are given a ``yes'' instance to the \textsc{X3C} problem; that is, there are $S_{i_1}, \hdots, S_{i_n}$ that disjointly cover $E$. Then, consider the following allocation:
    \begin{itemize}
        \item[---] In each category $\alpha_{i_j}$ corresponding to a set $S_{i_j}$, allocate to agent $s_{i_j}$ and the three element agents.
        \item[---] In the remaining $m-n$ set categories, allocate to four (distinct) filler agents arbitrarily.
        \item[---] In category $\beta$, allocate to agent $a$.
    \end{itemize}
    Note that this is a valid allocation. It satisfies \ax{\quota} and \ax{\eligibility} by construction, and it exhausts all quotas, so it is \ax{\pareto}. It  allocates to all filler and element agents, and to element agents only through categories whose set element is also allocated, so it is \ax{\priority}. This establishes that $a$ is serviceable. 

    Conversely, suppose that we reduce to an allocation instance in which $a$ is serviceable, so there is a valid integral (by \cref{lem:serv_int}) allocation $\varphi$ in which $\varphi(a) \neq \perp$. By construction, $a$ is eligible only in category $\beta$, so $a$ must receive the only unit of $\beta$. For $\varphi$ to respect priorities in $\beta$, it must allocate to each element agent, which must happen within the set categories. But then, to respect priorities in any set category, $\varphi$ must allocate to all filler agents as well. In total, these required allocations comprise $4m-n+1$ units, leaving $n$ units to allocate to the set agents $\{s_i\}$. Since the allocation to one set agent permits the allocation to at most three additional element agents, to allocate to all $3n$ element agents, $\varphi$ must allocate to \emph{exactly} $n$ set agents, and their corresponding sets must be pairwise disjoint. In summary, the $n$ set agents allocated in $\varphi$, $s_{i_1}, \hdots, s_{i_n}$, correspond to $n$ sets $S_{i_1}, \hdots, S_{i_n}$ that disjointly cover $E$, so we have reduced from a ``yes'' instance of \textsc{X3C}.
\end{proof}

In this context,~\citet{saban2015complexity} study the complexity of computing selection probabilities under \emph{random serial dictatorship} (where agents are ordered uniformly at random, and then pick their favorite remaining items in turn), and show that while one can efficiently identify items which have probability $1$ of being selected by some agent, it is \textsf{NP}-hard to identify items which have selection probability $0$. When specialized to this context, our results in this section recover and generalize this characterization.

%% file: s4b_utility.tex
\subsection{Incorporating Agent Utilities in Selecting Valid Allocations}
\label{sec:utility}

As a second extension, we consider how we can augment our basic model to incorporate agents' utilities for allocations from various categories. We relax our assumption of agent indifference, and equip each agent $a \in \A$ with a utility function $u_a : \C \to (0,1]$ that expresses the value that they derive from an allocation in each category. Thus, given an allocation $\x$, the \emph{realized utility} of agent $a \in A$ is given by
\[
    u_a(\x) = \sum_{c \in \C} u_a(c) \cdot x_{a,c}.
\]

In this setting, the natural objective is no longer only to allocate to as many agents as possible, but rather to use the realized utility of the agents to select valid allocations. There are two potential ways to do so: First, we can select valid allocations that are Pareto efficient with respect to agent utility. Alternatively, we can attempt to optimize some aggregate \emph{welfare} function of the agents' realized utilities. We next show that while the first goal admits an efficient algorithm, the second goal is \textsf{NP}-hard for most natural utility aggregation functions. 

First, we consider locating an allocation that is Pareto efficient with respect to agent utilities. To this end, in this section, we denote our usual notion of \emph{category-side} Pareto efficiency (i.e., the \ax{\pareto} property defined in~\cref{sec:model}) by \ax{C-\pareto}, and formalize Pareto efficiency from the viewpoint of agents as follows.

\begin{tcolorbox}
\begin{description}
    \item[\ax{A-\pareto} Agent-side Pareto Efficient:] An allocation $\x$ is \emph{Pareto efficient} with respect to agent utilities if there is no allocation $\y$ satisfying \ax{\quota}, \ax{\eligibility}, and \ax{\priority} such that:
    \begin{itemize}[nosep]
        \item[---] Each agent receives at least as much utility through $\y$: \quad $u_a(\y) \geq u_a(\x).$
        \item[---] At least one agent receives strictly higher utility through $\y$ than through $\x$.
    \end{itemize}
\end{description}
\end{tcolorbox}

Intuitively, an allocation is agent-side Pareto efficient if there is no incentive for the agents to attempt to trade their allocations (from the different categories); any trade would violate one of the other constraints (\ax{\eligibility} or \ax{\priority}), decrease some involved agent's utility, or leave all utilities unchanged. 
We now argue that we can select an agent-side Pareto efficient allocation via the following two-stage algorithm (\cref{alg:pe_matching_utilities}). At a high level, the first stage of the algorithm determines which agents will be included in the final allocation, while the second stage maximizes the utility realized by these agents.

\begin{algorithm}[H]
\caption{\textsc{Valid Allocation Selection with Agent-Side Pareto Efficiency}}  \label{alg:pe_matching_utilities}
\begin{algorithmic}[1]
    \STATE \textbf{Input:} Allocation instance $\mathcal{I}$ and agent utilities $( u_a )_{a \in \A}$
    \STATE Solve $(P_\delta)$ for any valid $\delta$ to pick an integral valid allocation $\x$
    \STATE $A \gets \big\{ a \in \A : \sum_{c} x_{a,c} = 0 \big\}$ 
    \STATE $\big( \C, \A, (q_c), (\E'_c), (\succeq'_c) \big) \gets \mathcal{I}_{\setminus A}$
    \STATE Define $U = \max\limits_{a,c}\{u_{a,c}\}$ and $\delta'_{a,c} = \frac{U - u_a(c)}{2|\A|\,|\C|}$ \quad (Note: higher $u_a(c) \implies$ smaller $\delta'_{a,c}$) 
    \STATE Solve $(P_{\delta'})$ for instance $\mathcal{I}_{\setminus A}$ to locate an integral allocation $\y$ \\
    \STATE \textbf{Return:} allocation corresponding to $\y$
\end{algorithmic}
\end{algorithm}

\begin{theorem} 
\label{thm:utility_pe_2step}
    \cref{alg:pe_matching_utilities} computes an allocation satisfying \emph{\ax{\quota}}, \emph{\ax{\eligibility}}, \emph{\ax{\priority}}, and \emph{\ax{A-\pareto}}.
\end{theorem}

\begin{proof}
    The constraints of $(P_{\delta'})$ ensure that $\y$ satisfies \ax{\quota} and \ax{\eligibility} in $\mathcal{I}_{\setminus A}$. Moreover, since the restriction operation leaves quotas unchanged and reduces the set of eligible agents, this allocation also satisfies these desiderata in $\mathcal{I}$. 
    
    Next, note that $\y$ is a \ax{\quota}, \ax{\eligibility}, \ax{\priority}, and maximal allocation in $\mathcal{I}_{\setminus A}$. By construction,
    \[
        \sum_{c \in \C} \sum_{a \in \A} \delta'_{a,c} \cdot x_{a,c}  
        \leq
        \sum_{c \in \C} \sum_{a \in \E'_c} \delta'_{a,c} 
        \leq
        \sum_{c \in \C} \sum_{a \in \E'_c} \tfrac{1}{2|\A|\,|\C|}
        \leq 
        \tfrac{1}{2}.
    \]
    Therefore, the objective values satisfy
        \[
            V(\x) \geq V(\y) 
            \geq V_{\delta'}(\y) 
            \geq V_{\delta'}(\x) 
            = V(\x) - \sum_{c \in \C} \sum_{a \in \A} \delta'_{a,c} \cdot x_{a,c}
            \geq V(\x) - \tfrac{1}{2}.
        \]
    Using the fact that $\x$ and $\y$ are integral allocations, we find that $V(\x) = V(\y)$. Thus, $\y$ allocates to all agents in the restricted instance, so it satisfies \ax{\priority}; the definition of the restriction $\mathcal{I}_{\setminus A}$ ensures that no unallocated agent in $A$ has priority over an agent allocated in $\y$.

    It remains to argue that $\y$ is agent-side Pareto efficient. The perturbations $\delta'_{a,c}$ are monotone decreasing in the utilities $u_a(c)$. Therefore, $\y$ maximizes the total utility of the allocated agents within $\mathcal{I}_{\setminus A}$. Therefore, any alternate allocation in which one agent realizes a higher utility must also include an agent who realizes a lower utility, so the allocation given by $\y$ satisfies \ax{A-\pareto}.
\end{proof}

Note that \cref{alg:pe_matching_utilities} does not admit an analogous realizability result to Theorem~\ref{thm:realizability}. As argued above, the computation of $\x$ in the first stage ensures that the final allocation is maximal. However, not every utility Pareto efficient allocation is maximal, as demonstrated by the example in \cref{fig:upe_not_max}.

\begin{figure}[H]
    \begin{center}
        \begin{minipage}{0.3\textwidth}
        \begin{center}
        $\x$: \quad
        \begin{tabular}{c|c}
             $\alpha$ {\scriptsize (1)} & $\beta$ {\scriptsize (1)} \\ \hline
             $a$ & $\boxed{a}$ \\
             $\boxed{b}$ & 
        \end{tabular}
        \end{center}
        \end{minipage}
        \quad
        \begin{minipage}{0.3\textwidth}
        \begin{center}
        $\y$: \quad
        \begin{tabular}{c|c}
             $\alpha$ {\scriptsize (1)} & $\beta$ {\scriptsize (1)} \\ \hline
             $\boxed{a}$ & $a$ \\
             $b$ & 
        \end{tabular}
        \end{center}
        \end{minipage}
    \end{center}    
    
    \caption{\small\it Consider the above instance with $u_{a,\alpha} = 1$, $u_{a,\beta} = u_{b,\alpha} = \frac{1}{3}$. The maximal (so \emph{\ax{C-\pareto}}) allocation shown on the left is \emph{\ax{A-\pareto}}, and will be output by \cref{alg:pe_matching_utilities}. However, the allocation shown on the right is also \emph{\ax{A-\pareto}}, and moreover utility-maximizing. However, as this allocation is not \emph{\ax{C-\pareto}}, it cannot be realized by \cref{alg:pe_matching_utilities}.}
    \label{fig:upe_not_max}
\end{figure}

Next, we turn our attention to the hardness of maximizing aggregate functions of the agents' realized utilities. Our main result in this setting is captured by the following theorem:

\begin{theorem}
\label{thm:max_utility_hard}
    Let $(F_n: [0,1]^n \to \mathbb{R})_{n=1}^{\infty}$ be a family of aggregation functions that are all continuous and strictly increasing in each of their arguments. Then, the following problem is NP-hard: Given an allocation instance $\mathcal{I}$ with $\A = \{ a_1, \hdots, a_n \}$, select a valid allocation maximizing the aggregate agent utility under $F_n$, i.e., find 
    \[
        \x^* \in \argmax_{\x \; \emph{valid}} \left\{ F_n \Big( u_{a_1}(\x) \;,\; \hdots \;,\; u_{a_n}(\x) \Big) \right\}.
    \]     
\end{theorem}

\begin{proof}
    We restrict attention to utilities of the following form:
    \begin{align*}
        u_{a_1}(c) &= 1 \quad \textrm{ for all } c \in \C \\
        u_{a_j}(c) &= u \quad \textrm{ for all } c \in \C \qquad \textrm{ for all } j \geq 2
    \end{align*}
    for some $u \in (0,1]$; in other words, all agents are indifferent regarding which category they are allocated through, and there is (weakly) higher utility for allocating agent $a_1$. 
    We write $t_\x \in [0,1]^{|\A|}$ for the vector of total agent allocations with $(t_\x)_j = \sum_{c} x_{a_j,c}$. 
    We then define
    \begin{align*}
        f & : (0,1] \times [0,1]^n \to \mathbb{R}
        & f \left( u, t_\x \right) 
        & = F_n \left( (t_\x)_1, u \cdot (t_\x)_2, \hdots, u \cdot (t_\x)_n \right)
    \end{align*}
    as the aggregate agent utility for a given parameter $u$ and agent allocation $\x$. Inheriting properties of $F_n$, $f$ is continuous and strictly increasing in $u$, and strictly increasing in each $(t_\x)_j$ when $u > 0$. 

    Now, suppose that there is some valid allocation $\x$ with $(t_\x)_1 = \tau > 0$ (i.e., $\x$ allocates to $a_1$) and another valid allocation $\y$ with $(t_{\y})_1 = 0$ (i.e., $\y$ does not allocate to $a_1$). Let $\mathbf{e}_1$ denote the first standard basis vector and $\mathbf{1}$ denote the all ones vector, both in $\mathbb{R}^{|\A|}$. 
    Since $f(0, \tau \cdot \mathbf{e}_1) > f(0, \mathbf{0}) = f(0,\mathbf{1}-\mathbf{e}_1)$ and $f$ is continuous in its first argument, then we can choose some sufficiently small $\varepsilon > 0$ such that
    \[
        f(\varepsilon, \tau \cdot \mathbf{e}_1) > f(\varepsilon, \mathbf{1}-\mathbf{e}_1).
    \]
 Since $f$ is strictly increasing in each agent's total allocation, the two allocations $\x,\y$ have aggregate utilities
    \[
    f( \varepsilon, t_\y ) 
    \leq f(\varepsilon, \mathbf{1} - \mathbf{e}_1) 
    < f(\varepsilon, \tau \cdot \mathbf{e}_1) 
    \leq f( \varepsilon, t_\x ).
\]
Therefore, we can reduce the problem of deciding whether $a_1$ is serviceable to determining whether the $F_n$-maximizing valid allocation $\x^*$ (under the utilities defined above) has value greater than $f(\varepsilon, \mathbf{1} - \mathbf{e}_1)$. 
From~\cref{prop:servicible_hard} we know that checking whether an agent is serviceable is \textsf{NP}-hard --- hence, so is the problem of selecting a valid allocation that maximizes aggregate utility.
\end{proof}

Note that this theorem relies on the fact that agents have \textit{cardinal} utilities for categories. We can use this theorem to conclude that many natural welfare optimization problems are computationally hard in the reserve allocation setting. We record two such results below. 

\begin{corollary}
    Given an instance $\mathcal{I}$ equipped with a utility function $u_a$ for each agent $a \in \A$, it is \textsf{NP}-hard to find a valid allocation $\x^*$ that maximizes total agent utility 
    \[
        \sum_{a \in \A} \sum_{c \in \C} u_a(c) \cdot x_{a,c}.
    \] 
\end{corollary}

This corollary provides a stark contrast to our original setting (concerned with the number of allocated agents), where maximizing total allocation and ensuring Pareto efficiency were equivalent (see~\cref{prop:pe_is_max}).

\begin{corollary}
    Given an instance $\mathcal{I}$ equipped with a utility function $u_a$ for each agent $a \in \A$, it is \textsf{NP}-hard to find a valid allocation $\x^*$ that maximizes Nash social welfare
    \[
        \emph{NSW}(\x) := \bigg( \prod_{a \in \A} \sum_{c \in \C} u_a(c) \cdot x_{a,c} \bigg)^\frac{1}{|\A|}.
    \]
\end{corollary}

%% file: s4c_threshold.tex
\subsection{Auditing Valid Allocations via Optimizing Cutoffs}
\label{sec:thresholds}

Thus far, we have considered the quality of allocations only through the formal desiderata that we have introduced. While theoretically satisfying, such an approach fails to acknowledge their impact on agents affected by these algorithms in practice. How can we convince the recipients (or, more aptly, non-recipients) of medical care, school seats, or other resources that decisions have been made fairly? This is discussed in great detail by \citet{pathak2021fair}, who suggest that one way addressing this issue is via the notion of \emph{auditability}: revealing extra information to agents to help satisfy them that their allocation is appropriate. 
In particular, a natural way to audit allocations is by revealing allocation \emph{thresholds} (or \emph{cutoff vectors}~\citet{pathak2021fair}) in each category. 
In this section, we study how to select valid allocations to optimize some metric related to these thresholds.

For notational ease, throughout this case study, we restrict our attention to integral allocations
, realized as maps $\varphi: \A \to \C \cup \{\outside\}$ (where $\varphi(a)=c$ if and only if $x_{a,c}=1$, and $\varphi(a)=\outside$ corresponds to $a$ being unallocated, i.e., $\sum_c x_{a,c}=0$.).
This is natural for defining cutoffs, and also is without loss of generality since our approach in~\cref{thm:delta_perturb} naturally returns integral allocations. 

\begin{definition}[\bf Allocation Thresholds]
    Thresholds $\theta : \C \to \mathbb{N}$ corresponding to allocation $\varphi$ satisfy:
    \begin{itemize}[nosep]
        \item[---] Every agent allocated in category $c \in \C$ has rank equal to or less than $c$'s threshold, i.e.\\
        {\color{white}.}\hspace{3cm}$\varphi(a) = c \quad\implies\quad r_c(a) \leq \theta(c) \qquad \text{ for all } a \in \A.
        $
        \item[---] Every unallocated agent has rank equal to or greater than the threshold in each eligible category\\
        {\color{white}.}\hspace{1cm}
        $\varphi(a) = \outside \; \text{ and } \; a \in \E_c \quad\implies\quad r_c(a) \geq \theta(c) \qquad \text{ for all } a \in \A.$
    \end{itemize}
\end{definition}

There are two natural thresholds associated with any allocation $\varphi$ (see \cref{fig:thresholds} for a visualization):
\begin{itemize}[nosep]
    \item[---] The \textbf{inner threshold} of $\varphi$, denoted by $\underline{\theta}$, has $\underline{\theta}(c) = \max \{ r_c(a) : \varphi(a) = c \}$, the \emph{maximum rank over all agents allocated} in each category.
    \item[---] The \textbf{outer threshold} of $\varphi$, denoted by $\overline{\theta}$, has $\overline{\theta}(c) = \min \{ r_c(a) : \varphi(a) = \outside, a \in \E_c \}$, the \emph{minimum rank over all unallocated eligible agents} in each category. If all agents in category $c$ are allocated, we set $\overline{\theta}(c)$ equal to one more than the maximum eligible rank in the category.
\end{itemize}

\begin{figure}[H]
  \begin{minipage}[c]{0.3\textwidth}
  \centering  
    \begin{tabular}{c|c|c}
        $\alpha$ {\scriptsize (3)} & $\beta$ {\scriptsize (2)} & $\gamma$ {\scriptsize (2)} \\ \hline
        $\boxed{a_1}$ & $\boxed{a_5}$ & $\boxed{a_6}$ \\
        $\boxed{a_2}$ & \cellcolor{red!15} $\boxed{a_3} \;,\, a_8$ & $a_3$ \\
        $a_3$ & \cellcolor{red!15} $a_4$ & $a_1$ \\
        \cellcolor{red!15} $\boxed{a_4}$ & \cellcolor{red!15} $a_0 \;,\, a_9$ & \cellcolor{red!15} $\boxed{a_8}$ \\
        \cellcolor{red!15} $a_9$ & $a_1$ & \cellcolor{red!15} $a_7$ \\
        $a_8$ & & $a_0$ \\
    \end{tabular}
  \end{minipage}\hfill
  \begin{minipage}[c]{0.65\textwidth}
    \caption{\small\it In this allocation instance (where the boxed agents form a valid allocation), the inner threshold $\underline{\theta}=(4,2,4)$ corresponds to the rank of the highest red-shaded tier in each column; all allocated agents occur at that priority level or higher. The lowest red-shaded tier corresponds to the outer threshold $\overline{\theta}=(5,4,5)$; the three unallocated agents ($a_0$, $a_7$ and $a_9$) are at or below this level in each category. Any mapping from the categories to one of the red-shaded tiers gives a valid threshold function.} 
    \label{fig:thresholds}
  \end{minipage}
\end{figure}



\noindent\textbf{Auditing Allocated Agents by Optimizing Inner Thresholds:}
One way to audit a valid allocation is by the quality of \emph{allocated} agents.
Allocations with large inner threshold are the ``most'' respectful of priorities in the sense that each category allocates only to agents in high priority tiers.
There are two natural ways to quantify this: we can minimize the \emph{sum} of ranks of allocated agents, or 
we can minimize the maximum rank of an allocated agent.
Both of these objectives are handled by our approach by carefully choosing the valid perturbation $\delta$.  

\begin{proposition}
\label{prop:innerthrsum}
    Given an instance $\mathcal{I}$, define perturbations $\delta_{a,c} = \frac{r_c(a)}{2|\C|\,{|\A|}^2}$. Then any (integral) allocation $\x$ returned by $(P_\delta)$ is a valid allocation that minimizes the sum of allocated agents' ranks. 
\end{proposition}

\begin{proof}
    To see that $\x$ is a valid allocation, it suffices (by Theorem~\ref{thm:delta_perturb}) to argue that $\delta$ is a valid perturbation. By construction, each $\delta_{a,c}$ is positive, and $\delta$ is consistent as $r_c(a) \leq r_c(a')$ if and only if $a \succeq_c a'$. Finally, to see that $\delta^A$ has small effect, note that each $r_c(a) \leq |\A|$, and hence
    \[
        \sum_{a \in \A} \sum_{c \in \C} \delta_{a,c} \leq \sum_{a \in \A} \sum_{c \in \C} \tfrac{1}{2|\C|\,|\A|} = \tfrac{1}{2}.
    \]
    To conclude that $\x$ minimizes the sum of allocated agents' ranks (among all valid allocations), we consider the objective $V_\delta(\x)$. We have
    \[
        V_{\delta}(\x) 
        = V(\x) - \sum_{a \in \A} \sum_{c \in \C} \delta_{a,c} \cdot x_{a,c} 
        = V(\x) - \tfrac{1}{2|\C|\,{|\A|}^2} \cdot 
        \Big( \sum_{a \in \A} \sum_{c \in \C} r_c(a) \cdot x_{a,c} \Big).
    \]
    $V(\x)$ is the same for all valid (and thus maximal) allocations. The parenthesized expression is exactly the sum of allocated agents' ranks. Thus, allocations returned by $(P_\delta)$  minimize this sum.
\end{proof}

\begin{proposition}
\label{prop:innerthrminmax}
    Given an instance $\mathcal{I}$, define perturbations $\delta_{a,c} = \frac{1}{2|\C|\,|\A|} \cdot \left( \frac{1}{|\A|+1} \right)^{|\A|-r_c(a)}$. Then any (integral) allocation $\x$ returned by $(P_\delta)$ is a valid allocation that minimizes the maximum rank over all allocated agents (i.e., maximum inner threshold over all categories).
\end{proposition}

\begin{proof}
    To see that $\x$ is a valid allocation, it suffices (by~\cref{thm:delta_perturb}) to argue that $\delta$ is a valid perturbation. As before, by construction, each $\delta_{a,c}$ is positive, and $\delta$ is consistent as $r_c(a) \leq r_c(a')$ if and only if $a \succeq_c a'$, and $\delta_{a,c}$ is an increasing function in $r_c(a)$. Finally, $\delta$ has small effect since each $r_c(a) \leq |\A|$, and so we have that $\big( \frac{1}{|\A|+1} \big)^{|\A|-r_c(a)} \leq 1$. Thus,  
    \[
        \sum_{a \in \A} \sum_{c \in \C} \delta_{a,c} \leq \sum_{a \in \A} \sum_{c \in \C} \tfrac{1}{2|\C|\,|\A|} = \tfrac{1}{2}.
    \]

    Let $R(\x) = \underset{(a,c) : x_{a,c} = 1}{\max} \big\{ r_c(a) \big\}$ be the maximum rank over all allocated agents. To conclude that $\x$ minimizes $R(\x)$ (among all valid allocations), consider the objective $V_\delta(\x)$. We have
    \[
        \notag V_{\delta}(\x) 
        = V(\x) - \sum_{a \in \A} \sum_{c \in \C} \delta_{a,c} \cdot x_{a,c} \\
        = V(\x) - \frac{1}{2|\C|\,|\A|\cdot (|\A|+1)^{|\A|}} \cdot \sum_{(a,c) : x_{a,c} = 1} (|\A|+1)^{r_c(a)}.
    \]
    By the definition of $R(\x)$, the sum in the last expression falls in the interval $\Big[ (|\A|+1)^{R(\x)}, |\A| \cdot (|\A|+1)^{R(\x)} \Big]$. Since these intervals are non-overlapping, choosing an integral allocation maximizing $V_{\delta}$ is equivalent to minimizing this sum, and hence minimizing $R(\x)$.
\end{proof}

\noindent\textbf{Auditing Unallocated Agents by Optimizing Outer Thresholds:}
Suppose instead that from the perspective of categories, what matters is that highly-ranked agents are allocated from \emph{some} category. 
A natural way to audit this is via the \emph{outer threshold}, which marks the rank of the first unallocated agent in a category; one may thus want to select valid allocations that have larger values for these outer thresholds. 
Again, there are two natural realizations of this objective: we can maximize the \emph{minimum} outer threshold, or the \emph{sum} of the outer thresholds over categories. Unlike the inner threshold, however, optimizing both of these objectives is \textsf{NP}-hard. 

\begin{proposition}
\label{prop:outerthrmaxmin}
Given an instance $\mathcal{I}$, selecting a valid allocation $\varphi^*$ that maximizes the minimum over all categories of the outer threshold is \textsf{NP}-hard.
\end{proposition}

\begin{proof}
    This result follows from an \textsc{X3C} reduction that is similar to that from~\cref{prop:servicible_hard}. In particular, when $4(m-n) \geq 3n+1$, the same reduction works, as $a$ is serviceable in the reduced instance if and only if the outer threshold of all categories is at least $3n+1$. 
    
    If $4(m-n) < 3n+1$, we need to add more filler agents to the $\alpha$ categories to push the tier of the last $f$ agents past the $a$ agent in category $\beta$. In category $\alpha_i$, we add agents $g_{i,1}, \hdots, g_{i,(7n-4m+1)}$, each in a separate rank tier above $f_1$. We also increase the category's quota to $7n-4m+5$. Again, we have that $a$ is serviceable in the reduced instance (so the $\textsc{X3C}$ instance has a partition by a straightforward modification of the proof of \cref{prop:servicible_hard} to account for the $g_{i,j}$ agents) if and only if the outer threshold of all categories is at least $3n+1$.
\end{proof}

\begin{proposition}
\label{prop:outerthrsum}
    Given an instance $\mathcal{I}$, selecting a valid allocation $\varphi^*$ that maximizes the sum over all categories of the outer threshold is \textsf{NP}-hard.
\end{proposition}

\begin{proof}
    This result again follows from an \textsc{X3C} reduction that is almost identical to that from~\cref{prop:servicible_hard}. To the reduced instance, we add $4m$ additional filler agents $\{g_1, \hdots, g_{4m}\}$ and an additional category $\gamma$ with quota $4m$ and all of these $g$ agents in its first priority tier. In addition, we add all of these $g$ agents below $a$ in category $\beta$, each in a separate priority tier. 

    Note that if $a$ remains unallocated, then the maximum possible sum of outer thresholds is $(4m-4n+5)\cdot m + 3n + 1$, where the first term comes from the $m$ set categories, the second term comes from $\beta$, and the third term from $\gamma$. On the other hand, if $a$ is allocated, then the sum of outer thresholds is at least $(4m-4n+1)\cdot m + (3n+1+4m) + 1$. Thus, $a$ is serviceable in the reduced instance (so the $\textsc{X3C}$ instance has a partition by the proof of~\cref{prop:servicible_hard}) if and only if the sum over categories of the outer thresholds is at least $(4m-4n+1)\cdot m + (3n+1+4m) + 1$.
\end{proof}

More surprisingly, there is a sense in which the second objective is strictly harder: suppose we \emph{de-reserve} units from the categories by removing the quota constraints, and instead impose a single global constraint that the total number of allocations across all categories is at most $q$. Now, maximizing the first objective becomes trivial (one can iteratively assign to the highest-ranked unallocated agent over all categories), but the objective of maximizing the sum of outer thresholds remains hard.
\begin{proposition}
\label{prop:outerthrsumgen}
Given an instance $\mathcal{I}$, selecting an allocation $\varphi^*$ giving to at most $q$ agents that maximizes the sum over all categories of the outer threshold is \textsf{NP}-hard.
\end{proposition}

\begin{proof}
    We perform a reduction from $\textsc{Clique}$. Given an undirected graph $G$ and clique size $k$ as input, construct an allocation instance $\mathcal{I}_G$ with $\A = V$, $q = k$, and a category $c_e$ for each edge $e \in E$ whose only two eligible agents are the endpoints of $e$ (in the same priority tier).  Now $G$ contains a $k$-clique if and only if the sum of outer thresholds in $\mathcal{I}_G$ equals $\binom{k}{2} + |E|$. 
\end{proof}

%% file: s5_online.tex
\section{Online Priority-Respecting Allocation}
\label{sec:online}

The second broad application we consider is allocating resources to agents who arrive \emph{online}, while still respecting priority and quota considerations.
Our results here again critically depend on the equivalence between valid matchings and perturbed maximum-weight matchings, demonstrating the importance of our characterization in~\cref{thm:delta_perturb}.

\subsection{Online Allocation with Priorities: Preliminaries}
\label{ssec:onlinemodel}

Our model is as follows:
Agents arrive one at a time over $T$ rounds $t=1, \ldots, T$; we refer to the agent arriving in round $t$ simply as \emph{agent $t$}.
Each arriving agent has an observable \emph{type} $\theta[t] \in \Theta$; here, $\Theta$ is a discrete and typically small set.
For example, each category could give each agent a \emph{priority level} in $\{1,2,\ldots,\ell,\text{ineligible}\}$ for some small $\ell$; in this case, an agent's type is their vector of priority levels.
Categories now have eligibility criteria and priorities over these types; that is, the eligible set is $\E_c \subseteq \Theta$, and the total pre-order $\preceq_c$ is defined over $\E_c$.
By distinguishing between agents and their types, our model allows us to separate out two parameters: the number of types (which is typically small), and the number of agents (which may be large).
Indeed, our main goal is to achieve online algorithms whose costs can be bounded in terms of the ``small'' parameters (number of types and number of categories), independent of the total number of agents $T$.

In each round $t$, the type $\theta[t]$ of the arriving agent is drawn randomly from some known probability distribution; for simplicity\footnote{Under suitable technical assumptions, our results can be generalized to non-stationary arrival probabilities. However, the added notational overhead outweighs the mild added generalization.}, we assume that $\theta[t]=\theta$ i.i.d.~with probability $p_\theta$. We use $\p=(p_\theta)_{\theta \in \Theta}$ for the vector of all these probabilities.
Under this arrival model, the number of agents $(N_\theta)_{\theta\in\Theta}$ of each type on a given sample path follows a $\text{Multinomial}(T,\p)$ distribution.

After observing the type $\theta[t]$ of agent $t$, the principal must irrevocably decide to either allocate a reserved unit from one of the categories to agent $t$ or leave $t$ unallocated forever.
Given the online nature of the problem and uncertainty due to randomness, it is impossible to satisfy all of the axioms we considered earlier; Pareto efficiency stands in obvious conflict with respecting priorities. To see this, notice that when an algorithm early on considers allocating to an agent of a particular type, there are two possible extreme scenarios that could occur with positive probability: if all subsequent agents have lower priority, then not allocating to the agent may result in a drastic loss in efficiency. Conversely, if all subsequent agents have higher priority, then allocating to the agent would deprive one of the future agents of an allocation, violating priorities. Thus, it is important to decide how to quantitatively trade off the violated axioms.

One natural option is to treat the priorities as a hard constraint, and maximize the expected number of allocations subject to this constraint.
Doing so leads to a straightforward MDP; unfortunately, treating priorities as a hard constraint can lead to very poor performance.

\begin{proposition}
\label{prop:onlinebad}
    Even with a single category and three priority levels (types), there exist instances in which any online allocation algorithm guaranteeing no priority violations must incur $\Omega(T)$ efficiency loss in hindsight, with all but at most exponentially small probability.
\end{proposition}

\begin{proof}{\cref{prop:onlinebad}}
    We consider a family of allocation instances
    parameterized by $T$.
    There is a single category with quota $q = \frac{T}{2}$ and three eligible types: $a \succ b \succ c$ with $p_a = p_b = p_c = \frac{1}{3}$. 

    Consider the arrival of an agent of type $b$ ``early in the sequence''. Although it is almost certain that the hindsight-optimal allocation will accept roughly half of the arriving type-$b$ agents, an online algorithm must reject this agent. To ensure that the hindsight allocation \emph{always} respects priorities, the algorithm must guard against a future (which occurs with positive probability) in which all remaining agents are of type $a$.
    The algorithm can therefore never exhaust the quota in a way that would leave some of these agents unallocated and envious.
    By this reasoning, the algorithm must continue to reject all arriving agents of types $b$ and $c$ until the (random) stopping time $\tau$ at which
    \[
        \tfrac{T}{2} - \sum_{t=1}^{\tau} \mathbbm{1} \big( \theta_t = a \big) \:\geq\: T-\tau.
    \]
    Here, the left-hand side is the number of remaining units that can be allocated, and the right-hand side is the number of agents to arrive after time $\tau$. We can rearrange this inequality to get 
    \[
         \sum_{t=1}^{\tau} \mathbbm{1} \big( \theta_t \ne a \big) \:\geq\: \tfrac{T}{2}.
    \]
    If at most $\frac{3T}{8}$ agents among the first $\frac{7T}{8}$ arrivals have type $a$, then this inequality holds for $\tau = \frac{7T}{8}$, so applying Hoeffding's Inequality, we obtain that
    \[
      \mathbb{P} [\tau \leq \tfrac{7T}{8}]
      \geq 1 - \mathbb{P}\big[ \text{Binom}(\tfrac{7T}{8}, \tfrac{1}{3}) > \tfrac{3T}{8}\big]
      \geq 1 - \exp\big(\tfrac{-T}{63}\big).
    \]
    After arrival $\tau$, an algorithm can begin to accept agents of type $b$ (and possibly $c$).
    However, if the algorithm has rejected any agents of type $b$ before time $\tau$, it cannot accept any type-$c$ agents.
    Since all agents of types other than $a$ must have been rejected before time $\tau$, the event that no type-$b$ agents have been rejected before time $\tau$ coincides with the event that no such agents arrived.
    Because $\tau \geq \frac{T}{2}$, the probability of no type-$b$ rejections is therefore at most $(\frac{2}{3})^{\nicefrac{T}{2}}$.
    By a union bound, with probability at least 
    \[
      1 - \exp\big(\tfrac{-T}{63}\big) - \big(\tfrac{2}{3}\big)^{-\nicefrac{T}{2}}
      = 1 - \exp(-\Omega(T)),
    \]
    the algorithm rejects all type-$c$ agents arriving after time $\tau \leq \frac{7T}{8}$. Again by Hoeffding's Inequality, with probability at least $1-\exp(-\Omega(T))$, there are at least $\big(\frac{1}{24} - \epsilon\big) \cdot T$ such agents (for any constant $\epsilon < \frac{1}{24}$, e.g., $\epsilon = \frac{1}{100}$), resulting in $\Omega(T)$ loss in efficiency with all but exponentially small probability.
\end{proof}

In light of \cref{prop:onlinebad}, it is necessary to relax the \ax{\priority} axiom to achieve meaningful guarantees. We therefore consider the tradeoff between the following two metrics:
\begin{tcolorbox}
\textbf{Efficiency loss} $\Delta_e$: The difference between the maximum cardinality of any allocation and the number of allocations made by the algorithm.\\
\textbf{Priority loss} $\Delta_p$: The number of unallocated agents with some type $\theta$ eligible in some category $c$ that allocated one or more slots to lower-priority agents (i.e., with type $\theta' \prec_c \theta$).
\end{tcolorbox}
We henceforth refer to unallocated agents contributing to the priority loss as \emph{priority violations}.
Note that both $\Delta_e$ and $\Delta_p$ are random variables, computed in hindsight on each sample path.
Moreover, the optimal \emph{offline} (i.e., hindsight) allocation simultaneously makes both losses $0$.
Our goal is to understand how online algorithms can trade off between these losses.

\subsection{Efficiency-Priority Tradeoffs for Online Allocation}
\label{ssec:onlinealgo}

We now present our main result in this section: we design an online allocation policy that guarantees that \emph{the sum of the efficiency loss and priority loss is independent of \,$T$ \!and $q$} (i.e., of the number of agents/allocations). Formally, we have the following guarantee.
\begin{theorem}
\label{thm:online}
  Let $\pmin = \min\limits_{\theta\in\Theta} p_{\theta}$.
  For any valid $\delta$, the allocation returned by the \emph{\textsc{Online Priority-Respecting Allocation with Restrictions}} Policy (\cref{alg:online}) satisfies
\begin{align*}
\mathbb{E}[\Delta_e + \Delta_p] & \leq \tfrac{|\Theta|^5(|C|+1)^4}{\pmin^4}.       
\end{align*}
\end{theorem}

The dependence of this bound on each of these three parameters ($\Theta$, $\C$, and $\pmin$) is unavoidable. Note that given any problem instance, we can duplicate each category with distinct types in each copy, leading to at least linear dependence on $|C|$ and $|\Theta|$. In addition, note that the problem of selecting the top $k$ elements of a random stream is a special case of our setting. For this problem, it is known that linear dependence on $\tfrac{1}{p_{\min}}$ is unavoidable (see Figure 1 in \cite{arlotto2019uniformly}). Getting the optimal dependence on $|\Theta|, p_{\min}$ and $|C|$ is left open for future work.

The central idea behind our algorithm is to solve the perturbed LP on the expected number of future arrivals and use the solution to select an action that is least likely to cause priority violations or efficiency loss. The guarantee follows by using the compensated coupling technique of~\citet{vera2021bayesian} (see also~\citet{banerjee2020uniform}), which essentially allows us to leverage smoothness properties of linear programs to obtain sample-path regret bounds. Our characterization in~\cref{thm:delta_perturb} is essential for using this approach. We note also that since our objective (in particular, $\Delta_p$) has a Lipschitz constant that grows with $T$, we cannot directly adopt existing uniform-regret results~\cite{banerjee2020uniform}; rather, we must carefully use \emph{restrictions} (\cref{def:restrict}) to control the Lipschitz constant and obtain our results.

Our algorithm uses as a subroutine the following
\textbf{Interim LP relaxation}    $P_\delta \Big(t, \N[t], \E[t], \q[t] \Big)$:

    \vspace{-10pt}
    \begin{align*}
        \max && \sum_{c \in \C} \sum_{\theta \in \Theta} x_{\theta,c}[t] \cdot (1 - \delta_{\theta,c}) \\
        \textrm{subject to} && x_{\theta,\outside}[t] + \sum_{c \in \C} x_{\theta,c}[t] &= N_\theta[t] & \textrm{for all $\theta \in \Theta$} \\
        && \sum_{\theta \in \Theta} x_{\theta,c}[t] &\leq q_c[t] & \textrm{ for all $c \in \C$} \\
        && x_{\theta,c}[t] &= 0 & \textrm{for all $c \in \C, \theta \not\in \E_c[t]$} \\
        && x_{\theta,c}[t] &\geq 0 & \textrm{for all $c \in \C \cup \{ \outside \}, \theta \in \Theta$}
    \end{align*}
The interim LP can be viewed as a proxy solution to the perturbed LP $(P_\delta)$ in~\cref{thm:delta_perturb}, \emph{given past allocation decisions}.
$t$ indexes the current arrival, the parameters $\N[t] = (N_\theta[t])_{\theta \in \Theta}$ represent the number of \emph{future} arrivals of each type $\theta \in \Theta$ over rounds $t, \ldots, T$, the parameters $\E[t] = (\E_c[t])_{c \in \C}$ represent the restricted eligibility sets (see~\cref{alg:online}) at time $t$, and the parameters $\q[t] = (q_c[t])_{c \in \C}$ represent the available quotas at the start of round $t$. 
The decision variables $\x[t]=(x_{\theta,c}[t])_{c \in \C \cup \{\outside\}, \theta \in \Theta}$ represent the number of agents of type $\theta$ who will be allocated in category $c$ (or remain unallocated, for $c = \outside$) from among the arrivals $t, \ldots, T$.
The objective function, as before, accrues one unit for each allocated agent minus some chosen perturbation $\delta_{\theta,c}$.
The first constraint accounts for future arrivals of type $\theta$; the second ensures that the combination of past and future allocations does not exceed the reserved quota for any category; the third ensures that the solution respects eligibility.
Note that the interim LP does not ensure respect for priorities, as it does not account for which agent types were allocated in the past. In fact, as shown in \cref{prop:servicible_hard}, it is \textsf{NP}-hard to compute whether there is a valid allocation that includes these agents.
Given this LP family, we are ready to state our algorithm.

\begin{algorithm}[ht]
    \caption{\textsc{Online Priority-Respecting Allocation with Restrictions}}
    \textbf{Input}: Allocation Instance $\big(\C, \Theta, \q, \E, (\preceq_c)_{c \in \C}, \p \big)$, Online Arrivals $(\theta[t])_{t \in [T]}$ \hfill \phantom{a} \\
    
    \textbf{Output}: Allocations $(y[t])_{t \in [T]}$, $y[t] \in \C \cup \{\outside\}$ \hfill
    \begin{algorithmic}[1] \label{alg:online}
        \STATE Select a valid perturbation 
        $\delta$;\quad Initialize \: $\E[1] \gets \E$, \quad  $\q[1]\gets \q$
        \FOR{each $t = 1, \ldots, T$}
            \STATE $\x^*[t]\gets$ solution to $P_\delta \Big(t, \big(\mathbbm{1}(\theta = \theta[t]) + (T-t) \cdot p_\theta\big)_{\theta \in \Theta},\E[t], \q[t]\Big)$
            \STATE $y[t] \gets \argmax\limits_{c \in \C \cup \{ \outside \}} \left( x^*_{\theta[t],c}[t] \right)$, \;\; $q_c[t+1]\gets q_c[t]-\mathbbm{1}(y[t]=c)$ for each $c \in \C$ 
            \IF{$y[t] = \outside$}
                \STATE $\E_c[t+1] \gets \E_c[t] \setminus \big( \{ \theta[t] \} \cup \{ \theta \in \Theta \colon \theta \prec_c \theta[t] \} \big)$ for each $c \in \C$
            \ELSE
                \STATE $\E_c[t+1] \gets \E_c[t]$ for each $c \in \C$
            \ENDIF
        \ENDFOR
    \end{algorithmic}
\end{algorithm}

For each arriving agent $t$, the algorithm solves the LP using its current quotas $\q[t]$ and eligible sets $\E[t]$, the current arrival $\theta[t]$, and the expected number of future arrivals of each type.
It allocates to agent $t$ through a category (including the ``no allocation'' category $\outside$) maximizing the expected allocation under the optimal LP solution.
When an agent is not allocated, the algorithm takes a restriction of the allocation instance to prevent future priority violations.  


\begin{proof}[Proof of~\cref{thm:online}]
  To bound the expected loss of \cref{alg:online}, we use a variant of the compensated coupling argument of~\citet{vera2021bayesian}. In each round $t$, we consider two random variables, with the randomness taken over the future arrivals $\theta[t+1], \ldots, \theta[T]$.
\begin{itemize}[leftmargin=0.5 cm]
    \item[---] $\Delta_e[t]$ represents the \emph{efficiency} loss due to the algorithm's decision at time $t$. Using our notation,
    \[
        \Delta_e[t] = \!\!\! \underbrace{P_0 \Big( t, \N[t],\E[t], \q[t] \Big)}_{\begin{matrix} \textrm{Optimal offline allocation given} \\ \textrm{decisions made before round $t$} \end{matrix}} \!\!\!\!\!-\:\:\:
        \Big( \mathbbm{1} ( y[t] \ne \outside) + \underbrace{P_0 \Big( t+1, \N[t+1],\E[t+1], \q[t+1] \Big)}_{\begin{matrix} \textrm{Optimal offline allocation given} \\ \textrm{decisions made through round $t$} \end{matrix}} \Big).
    \]
    \item[---] $\Delta_p[t]$ represents the \emph{priority} loss due to the algorithm's decision at time $t$, i.e., the number of additional unallocated and envious agents that arise as a result of the allocation of $\theta[t]$.
\end{itemize}

We denote the value of the decision variables at an optimum of the \emph{offline} LP at time $t$ by $\x^*[t]$.
We separately reason about these two sources of loss in two cases: when agent $t$ is allocated through some category, vs.~when $t$ remains unallocated.

If agent $t$ is allocated, then $y[t] = c$ for some $c \in \C$.
If $x^*_{\theta[t],c}[t] > 0$, then the optimal solution along this sample path allocates to an agent of type $\theta[t]$ in category $c$.
Thus, the allocation to agent $t$ has not deviated from this optimal allocation, so no loss needs to be compensated for.
If $x^*_{\theta[t],c}[t] = 0$, meaning that the optimal solution does not allocate to any agents of type $\theta[t]$ from time $t$ onwards, the algorithm's choice of allocation may reduce the efficiency by at most one.
This is because the optimal solution can introduce at most one augmenting path into the bipartite allocation graph.
In addition to the efficiency loss, the allocation to $t$  may prevent some agents with higher priority in $c$ from receiving an allocation, leading to priority violations.
A crude upper bound on the increase in the number of priority violations is $T-t$, i.e., all remaining agents. Hence, we obtain the upper bound 
$\Delta_e[t] + \Delta_p[t]  \leq \mathbbm{1}(x^*_{\theta[t],c}[t] = 0) \cdot (1 + T-t)$.  
  
Next, we consider the case in which agent $\theta[t]$ remains unallocated, so $y[t] = \outside$.
Again, if $x^*_{\theta[t],\outside}[t] > 0$, i.e., the optimum solution also leaves at least one agent of type $\theta[t]$ unallocated, the failure to allocate to agent $t$ does not cause any loss in efficiency or priority.
Therefore, we assume that $x^*_{\theta[t],\outside}[t] = 0$.
The non-allocation to $\theta[t]$ causes the algorithm to restrict the allocation instance: in the future, it will never be able to allocate to agents whose types have lower priority than $\theta[t]$.
Even so, the efficiency loss can be safely upper-bounded by $T-t+1$, i.e., all agents after and including agent $\theta[t]$.
In addition to the efficiency loss, the failure to allocate to agent $t$ may lead to a priority violation at the expense of $t$; however, this can be the only resulting priority violation.
Thus, we obtain the upper bound
$\Delta_e[t] + \Delta_p[t] \leq \mathbbm{1}(x^*_{\theta[t],\outside}[t] = 0) \cdot (T-t+1+1).$

Combining the above, we get that the sum of the losses in round $t$ can be upper-bounded as
\begin{align*}
    \Delta_e[t] + \Delta_p[t] & \leq \mathbbm{1} ( x^*_{\theta[t],y[t]}[t] = 0) \cdot (T-t+2), 
\end{align*}
and summing over all rounds, and taking expectations, we get
\begin{align} \label{eq:exploss}
    \mathbb{E}[ \Delta_e + \Delta_p] & \leq \sum_{t=1}^{T} \mathbb{P} [x^*_{\theta[t],y[t]}[t] = 0] \cdot (T-t+2).
\end{align}
Next, we establish a bound on the probability $\mathbb{P} [x^*_{\theta[t],y[t]}[t] = 0]$.
Recall that in each round, $y[t]$ is selected as an argmax over $c \in \C \cup \{\outside\}$ of $x_{\theta,c}[t]$.
That is, $y[t]$ is a most frequent assignment of the future arriving agents of type $\theta[t]$ when the expected number of agents of each type arrive.
In expectation, the number of arrivals of type $\theta[t]$ in rounds $t, \ldots, T$ is $1 + (T-t) \cdot p_{\theta[t]}$. Thus, $x_{\theta[t],y[t]} \geq \frac{1+(T-t) p_{\theta[t]}}{|\C|+1}$; this implies a lower bound on the infinity norm of the difference between the LP solution $\x[t]$ and the optimal offline solution $\x^*[t]$, i.e., $\big\|\x[t]-\x^*[t]\big\|_{\infty}\geq \frac{1+(T-t) \cdot p_{\theta[t]}}{|\C|+1}$.

On the other hand, using the $(1,\infty)$-Lipschitz property of maximum-weight matchings with respect to budgets~\cite[Proposition 4]{vera2021bayesian}, we have that
\begin{align*}
    \big\| \x^*[t] - \x[t] \big\|_\infty & \leq \big\| \N[t+1] - (T-t-1) \cdot \p \, \big\|_1.
\end{align*}

Thus, the (bad) event $x^*_{y[t],\theta[t]}[t] = 0$ implies that $\big\| \N[t+1] - (T-t-1) \cdot \p \, \big\|_1 \geq \frac{1+(T-t) \cdot p_{\theta[t]}}{|\C|+1}$, i.e., that the actual type counts differ a lot from their expectations.
A large deviation of the $\| \cdot \|_1$-norm implies that at least one coordinate must differ by at least the average, so this event implies that the actual number of arrivals for at least one type differs from its expectation by at least an additive $\frac{1+(T-t) \cdot p_{\theta[t]}}{ |\Theta| \cdot (|\C|+1)}$.
Because arrival counts for any type $\theta$ follow the distribution $N_{\theta}[t]\sim \textrm{Binomial}(T-t,p_{\theta})$, the Hoeffding bound gives us that the probability of a large deviation for any one type $\theta$ is
\begin{align*}
  \mathbb{P} \left[ \big| N_{\theta}[t+1] - \mathbb{E} \big[ N_{\theta}[t+1] \big] \big| \geq \tfrac{1+(T-t) \cdot p_{\theta}}{ |\Theta| \cdot (|\C|+1)} \right]
  & \leq 2 \cdot \exp \left( \tfrac{- 2(T-t) \cdot p^2_{\theta}}{|\Theta|^2 \cdot (|\C|+1)^2} \right)
  \; \leq \; 2e^{-\kappa (T-t)},
\end{align*}
where $\kappa = \frac{2\pmin^2}{|\Theta|^2 (|\C|+1)^2} \leq \frac{1}{2}$.
Taking a union bound over the $|\Theta|$ types $\theta$ and substituting the resulting upper bound into \eqref{eq:exploss}, the expected loss is upper-bounded by
\begin{align*}
\mathbb{E}[\Delta_e+\Delta_p]
  & \leq 
    \sum_{t=1}^{T} (T-t+2) \cdot 2 |\Theta| \cdot e^{-\kappa \cdot (T-t)} \\
    &\leq 2 |\Theta| \cdot \int_{0}^{\infty} \!\!\!(z+2) \cdot e^{-\kappa z} \;dz 
      \; = \; |\Theta| \cdot \tfrac{4\kappa + 2}{\kappa^2}
      \; \stackrel{\kappa \leq \frac{1}{2}}{\leq} \; |\Theta| \cdot \tfrac{4}{\kappa^2}
      \; = \; \tfrac{|\Theta|^5 (|\C|+1)^4}{{\pmin}^4}. \qedhere
\end{align*}
\end{proof}

%% file: s6_conclusion.tex
\section{Conclusion}
\label{sec:conclusion}

We studied allocation settings where units of some public resource are to be divided between multiple categories, each with a quota of items, and a priority ordering over eligible agents. The goal is to find a \emph{valid allocation} --- one which respects the quotas, eligibility, and priority requirements, while still being Pareto optimal. Our main result demonstrates a bijection between valid integral allocations and maximum-weight matchings under a set of \emph{valid weights}. This approach allowed us to efficiently locate and select valid allocations, despite the set of valid allocations being non-convex. On the other hand, our hardness results demonstrate the strange geometry of this set, due to which optimizing over it remains challenging.
We hope our work can help guide the use of priorities and quotas in a wide variety of settings. Extending our approach to models involving two-sided preferences and/or complementarities provide interesting avenues for future research.

%% file: a_scarf.tex
\section{Scarf's Lemma and Stable Matching}
\label{sec:scarf}

Here, we compare the priority-respecting allocations problem and the stable matching problem through the lens of Scarf's lemma. To begin, we recall Scarf's Lemma, following the treatment of \citet{nguyen2022complementarities}. 

Scarf's lemma considers an allocation setting with $n$ \emph{agents} and $m$ \emph{coalitions}; both of these descriptors apply rather abstractly, as our examples will illustrate. In this setting, coalitions comprise agents and the principal must decide how to allocate coalitions. There are budgetary constraints that ensure that no agent is over-allocated and agents may express preferences over the coalitions to which they belong. 
Formally, we have:
\begin{itemize}
    \item[---] A matrix $\mathbf{A} \in \mathbb{R}_{+}^{n \times m}$ has a row for each agent and a columns for each coalition.We interpret entry $\mathbf{A}_{ij}$ as a cost to agent $i$ if (one unit of) coalition $j$ is allocated. We assume that each row includes at least one positive entry.
    \item[---] A vector $\mathbf{q} \in \mathbb{R}_{+}^{n}$ denotes the budget of each agent.
    \item[---] Each agent $i$ has a total preference order $\succeq_i$ over its coalitions $\{ j \in [m] \colon \mathbf{A}_{ij} > 0 \}$.  
    \item[---] A vector $\mathbf{x} \in \mathbb{R}_{+}^{m}$ stipulates to what extent each coalition is realized. 
\end{itemize}

Thus, the principal must select $\mathbf{x}$ subject to the budgetary constraints $\mathbf{Ax} \leq \mathbf{q}$. Within this set of feasible $\mathbf{x}$, we wish to further choose coalitions that enforce some notion of stability with respect to the agent preferences. For this, we introduce the notion of the domination of a coalition. 

\begin{definition}
    Given an instance $\big(\mathbf{A}, \mathbf{q}, (\succeq_i)_{i \in [n]} \big)$, an allocation $\mathbf{x} \geq 0$ satisfying $\mathbf{Ax} \leq \mathbf{q}$ \emph{dominates} coalition $j \in [m]$ if there is some fully-allocated agent $i$ for which every allocated coalition to which $i$ belongs is weakly preferred by $i$ to $j$.
    
    More formally, there is $i \in [n]$ such that $\sum_{k=1}^{m} \mathbf{A}_{ik} \mathbf{x}_k = \mathbf{q}_i$ and for each $j' \in [m]$,
    \[
        \mathbf{A}_{ij'} > 0 \textrm{ and } \mathbf{x}_{j'} > 0 \implies j' \succeq_i j. 
    \]
\end{definition}

Domination expresses an inability to adjust $\mathbf{x}$ in a way that assigns more weight to coalition $j$ without upsetting some agent $i$. Simply increasing $\mathbf{x}_j$ would violate $i$'s budgetary constraint, and any shift in weight from any other coalition $j'$ to which $i$ belongs would come from a coalition preferable to $j$. Through this interpretation, if an allocation were to dominate all coalitions, it would exhibit a notion of \emph{stability}; any adjustment of the coalition allocations would be either inefficient or disagreeable to some agent. The ensured existence of such stable allocations is the content of Scarf's Lemma. 

\begin{proposition}[\citet{scarf1967core}, Theorem 1]
    Given any allocation instance $\big(\mathbf{A}, \mathbf{q}, (\succeq_i)_{i \in [n]} \big)$, there is an extreme point of $\{ \mathbf{x} \colon \mathbf{Ax} \leq \mathbf{q} \}$ that dominates every coalition.
\end{proposition}

Scarf's Lemma can be proven via a reduction to the existence of Nash Equilibria in two-person games. While the statement of this result is clean, allowing it to be specialized to many problems (as we discuss below), there is no assurance that this dominating extreme point can be easily computed. In fact, \citet{kintali2008complexity} showed that a computational version of Scarf's Lemma is complete for the \textsf{PPAD} class. This implies that there is no polynomial-time algorithm to locate these extreme points unless $\textsf{PPAD} \subseteq \textsf{P}$. Despite this, there are many special cases of Scarf's Lemma that admit polynomial algorithms. 

One special case of Scarf's Lemma is the stable matching problem; it can be used to recover the result of \citet{gale1962college} that a stable matching exists in every instance. Using the context of $n$ residents being matched to $n$ hospitals, we take the set of agents to be the union of the residents and hospitals. The coalitions consist of each (resident, hospital) pair, and $\mathbf{A}$ is the $\{0,1\}$ incidence matrix. The budget vector $\mathbf{q} \in \mathbb{R}_+^{2n}$ is the all-ones vector, which ensures that each agent is matched at most once. The agent rankings $(\succeq_i)$ order an agents incident pairs corresponding to their preference list. In this construction, an undominated coalition corresponds to an instability. Note that the Birkhoff-von Neumann theorem ensures the integrality of the extreme points. 



        

The priority-respecting allocation problem can also be interpreted as a special case of Scarf's lemma. However, the construction is less straightforward, as we must account for the lack of preferences of the agents. Here, the set of agents consists of all of the categories along with a copy of each of the agent for each possible ordering of their eligible categories. The set of coalitions consists of all eligible (agent, category) pairs, and $\mathbf{A}$ is again a $\{0,1\}$ incidence matrix. The budget of each category is its quota, and the budget of each agent is 1. The preference order of a category $c$ is any linear extension of $\succeq_c$. The preference order of each agent row corresponds to its ordering over its eligible categories.

\subsection{LP Perturbations and Stable Matchings.}

One problem with using Scarf's Lemma is that while it guarantees the existence of a dominating solution, it does not give an efficient algorithm for finding it. Apart from priority-respecting allocation, the other setting where the dominating solution was known to be efficiently computable was for stable matchings. The underlying reason behind the existence of an efficient algorithm in the two settings, however, appears to be very different. On one hand, stable matchings are known to form a convex set (and indeed, are realized as corner points of a natural modification of the matching LP~\cite{vate1989linear}), while as we show in~\cref{fig:non-convex}, this is not the case for priority-respecting matchings.
On the other hand, we show that the perturbation techniques we develop for locating priority-respecting allocations does not work for stable matchings.

A na\"{i}ve way to locate stable matchings via LPs is to first compute a stable matching (which can be done efficiently via the Deferred-Acceptance procedure of~\citet{gale1962college}), and then design edge weights to recover the same matching as a maximum-weight matching. More surprisingly, the work of Vande Vate~\citep{vate1989linear} shows that one can modify the matching polytope by adding additional linear constraints to get an LP whose corner points exactly correspond to all the stable matchings, and one can use the corresponding optimal dual variables to get perturbed objectives.  
The problem with these procedures, however, is that they compute perturbations that are \emph{global}, i.e., based on the entire instance. This is in contrast to our technique for finding priority-respecting allocations, which is based on \emph{local} perturbations: the objective coefficient on edges $x_{mw}$ are functions only of the rank of $m$ on $w$'s preference list and the rank of $w$ on $m$'s preference list. 

Thus, a more refined question is if given a stable matching instance with $n$ men and $n$ women, one can find a perturbation function $F \colon [n] \times [n] \to \mathbb{R}$ such that the resulting matching $M$ that maximizes $V_F(M) = \sum_{(m,w) \in M} F(r_w(m),r_m(w))$ is necessarily stable. Unfortunately, we can answer this question in the negative. 

\begin{proposition}
\label{prop:unstable}
    For $n\geq 6$, for any local perturbation function $F \colon [n] \times [n] \to \mathbb{R}$, there exist instances such that any matching $M$ maximizing $V_F(M) = \sum_{(m,w) \in M} F(r_w(m),r_m(w))$ is unstable.
\end{proposition}

\begin{proof}
    We consider two stable matching instances with $n=6$. In both instances, the women are indexed by Roman letters $\{a,b,c,d,e,f\}$ and the men are indexed by Greek letters $\{\alpha, \beta, \gamma, \delta, \epsilon, \zeta \}$. The first instance has the following preference lists:

    \vspace{10pt}
    \begin{minipage}{0.5\textwidth}
    \begin{center}
    \begin{tabular}{c|c|c|c|c|c}
        $a$ & $b$ & $c$ & $d$ & $e$ & $f$ \\ \hline 
        $\alpha$ & $\beta$ & $\gamma$ & $\alpha$ & $\alpha$ & $\alpha$ \\
        $*$ & $*$ & $*$ & $\delta$ & $\beta$ & $\epsilon$ \\
        $*$ & $*$ & $*$ & $\zeta$ & $\epsilon$ & $\beta$ \\
        $*$ & $*$ & $*$ & $*$ & $\delta$ & $\zeta$ \\
        $*$ & $*$ & $*$ & $*$ & $*$ & $*$ \\
        $*$ & $*$ & $*$ & $*$ & $*$ & $*$ \\
    \end{tabular}
    \end{center}
    \end{minipage}
    \begin{minipage}{0.5\textwidth}
    \begin{center}
    \begin{tabular}{c|c|c|c|c|c}
        $\alpha$ & $\beta$ & $\gamma$ & $\delta$ & $\epsilon$ & $\zeta$ \\ \hline 
        $a$ & $b$ & $c$ & $a$ & $a$ & $a$ \\
        $*$ & $*$ & $*$ & $e$ & $b$ & $f$ \\
        $*$ & $*$ & $*$ & $b$ & $e$ & $d$ \\
        $*$ & $*$ & $*$ & $c$ & $f$ & $*$ \\
        $*$ & $*$ & $*$ & $d$ & $*$ & $*$ \\
        $*$ & $*$ & $*$ & $*$ & $*$ & $*$ \\
    \end{tabular}
    \end{center}
    \end{minipage}

    \vspace{10pt}
    Here, the $*$ elements can be assigned arbitrarily to complete the matching instance. Note that in this instance, there is a unique stable matching, $M = \{ (a,\alpha), (b,\beta), (c,\gamma), (d,\delta), (e,\epsilon), (f,\zeta) \}$. An alternate (non-stable) matching is $M' = \{ (a,\alpha), (b,\beta), (c,\gamma). (d,\zeta), (e,\delta), (f,\epsilon) \}$; note the presence of instability $(e,\epsilon)$. For our function $F$ to assign a higher value to matching $M$ than $M'$, we must have
    \begin{align*}
        &V_F(M) = F(1,1) + F(1,1) + F(1,1) + F(2,5) + F(3,3) + F(4,2) \\
        > &V_F(M') = F(1,1) + F(1,1) + F(1,1) + F(3,3) + F(4,2) + F(2,4),
    \end{align*}
    which simplifies to the condition $F(2,5) < F(2,4)$. 

    We similarly consider our second stable matching instance. 

    \vspace{10pt}
    \begin{minipage}{0.5\textwidth}
    \begin{center}
    \begin{tabular}{c|c|c|c|c|c}
        $a$ & $b$ & $c$ & $d$ & $e$ & $f$ \\ \hline 
        $\alpha$ & $\beta$ & $\gamma$ & $\alpha$ & $\alpha$ & $\alpha$ \\
        $*$ & $*$ & $*$ & $\delta$ & $\beta$ & $\epsilon$ \\
        $*$ & $*$ & $*$ & $\zeta$ & $\epsilon$ & $\beta$ \\
        $*$ & $*$ & $*$ & $*$ & $\delta$ & $\zeta$ \\
        $*$ & $*$ & $*$ & $*$ & $*$ & $*$ \\
        $*$ & $*$ & $*$ & $*$ & $*$ & $*$ \\
    \end{tabular}
    \end{center}
    \end{minipage}
    \begin{minipage}{0.5\textwidth}
    \begin{center}
    \begin{tabular}{c|c|c|c|c|c}
        $\alpha$ & $\beta$ & $\gamma$ & $\delta$ & $\epsilon$ & $\zeta$ \\ \hline 
        $a$ & $b$ & $c$ & $a$ & $a$ & $a$ \\
        $*$ & $*$ & $*$ & $e$ & $b$ & $f$ \\
        $*$ & $*$ & $*$ & $b$ & $e$ & $d$ \\
        $*$ & $*$ & $*$ & $d$ & $c$ & $*$ \\
        $*$ & $*$ & $*$ & $*$ & $f$ & $*$ \\
        $*$ & $*$ & $*$ & $*$ & $*$ & $*$ \\
    \end{tabular}
    \end{center}
    \end{minipage}

    \vspace{10pt}
    In this instance, $M = \{ (a,\alpha), (b,\beta), (c,\gamma), (d,\delta), (e,\epsilon), (f,\zeta) \}$ is again the unique stable matching. The alternate matching $M' = \{ (a,\alpha), (b,\beta), (c,\gamma). (d,\zeta), (e,\delta), (f,\epsilon) \}$ is again unstable; note the presence of instability $(e,\epsilon)$. For our function $F$ to assign a higher value to matching $M$ than $M'$, we must have
    \begin{align*}
        &V_F(M) = F(1,1) + F(1,1) + F(1,1) + F(2,4) + F(3,3) + F(4,2) \\
        > &V_F(M') = F(1,1) + F(1,1) + F(1,1) + F(3,3) + F(4,2) + F(2,5),
    \end{align*}
    which simplifies to the condition $F(2,4) < F(2,5)$. Our two derived inequalities cannot be simultaneously satisfied. Hence, such a local perturbation function $F$ cannot exist.
\end{proof}